\documentclass{article}

\usepackage{amssymb,latexsym,amsmath}

\usepackage[pdftex]{graphicx}

\usepackage{hyperref}

\begin{document}

\pagestyle{plain}

\newtheorem{theorem}{Theorem}[section]

\newtheorem{proposition}[theorem]{Proposition}

\newtheorem{lema}[theorem]{Lemma}

\newtheorem{corollary}[theorem]{Corollary}

\newtheorem{definition}[theorem]{Definition}

\newtheorem{remark}[theorem]{Remark}

\newtheorem{exempl}{Example}[section]

\newenvironment{exemplu}{\begin{exempl}  \em}{\hfill $\square$

\end{exempl}}

\newcommand{\ea}{\mbox{{\bf a}}}

\newcommand{\eu}{\mbox{{\bf u}}}

\newcommand{\ueu}{\underline{\eu}}

\newcommand{\ueo}{\overline{u}}

\newcommand{\oeu}{\overline{\eu}}

\newcommand{\ew}{\mbox{{\bf w}}}

\newcommand{\ef}{\mbox{{\bf f}}}

\newcommand{\eF}{\mbox{{\bf F}}}

\newcommand{\eC}{\mbox{{\bf C}}}

\newcommand{\en}{\mbox{{\bf n}}}

\newcommand{\eT}{\mbox{{\bf T}}}

\newcommand{\eL}{\mbox{{\bf L}}}

\newcommand{\eR}{\mbox{{\bf R}}}

\newcommand{\eV}{\mbox{{\bf V}}}

\newcommand{\eU}{\mbox{{\bf U}}}

\newcommand{\ev}{\mbox{{\bf v}}}

\newcommand{\eve}{\mbox{{\bf e}}}

\newcommand{\uev}{\underline{\ev}}

\newcommand{\eY}{\mbox{{\bf Y}}}

\newcommand{\eK}{\mbox{{\bf K}}}

\newcommand{\eP}{\mbox{{\bf P}}}

\newcommand{\eS}{\mbox{{\bf S}}}

\newcommand{\eJ}{\mbox{{\bf J}}}

\newcommand{\eB}{\mbox{{\bf B}}}

\newcommand{\eH}{\mbox{{\bf H}}}

\newcommand{\leb}{\mathcal{ L}^{n}}

\newcommand{\eI}{\mathcal{ I}}

\newcommand{\eE}{\mathcal{ E}}

\newcommand{\hen}{\mathcal{H}^{n-1}}

\newcommand{\eBV}{\mbox{{\bf BV}}}

\newcommand{\eA}{\mbox{{\bf A}}}

\newcommand{\eSBV}{\mbox{{\bf SBV}}}

\newcommand{\eBD}{\mbox{{\bf BD}}}

\newcommand{\eSBD}{\mbox{{\bf SBD}}}

\newcommand{\ecs}{\mbox{{\bf X}}}

\newcommand{\eg}{\mbox{{\bf g}}}

\newcommand{\paromega}{\partial \Omega}

\newcommand{\gau}{\Gamma_{u}}

\newcommand{\gaf}{\Gamma_{f}}

\newcommand{\sig}{{\bf \sigma}}

\newcommand{\gac}{\Gamma_{\mbox{{\bf c}}}}

\newcommand{\deu}{\dot{\eu}}

\newcommand{\dueu}{\underline{\deu}}

\newcommand{\dev}{\dot{\ev}}

\newcommand{\duev}{\underline{\dev}}

\newcommand{\weak}{\stackrel{w}{\approx}}

\newcommand{\mild}{\stackrel{m}{\approx}}

\newcommand{\lrightarrow}{\stackrel{L}{\rightarrow}}

\newcommand{\rrightarrow}{\stackrel{R}{\rightarrow}}

\newcommand{\strong}{\stackrel{s}{\approx}}

\newcommand{\weakdown}{\rightharpoondown}

\newcommand{\opg}{\stackrel{\mathfrak{g}}{\cdot}}

\newcommand{\opunu}{\stackrel{1}{\cdot}}
\newcommand{\opdoi}{\stackrel{2}{\cdot}}

\newcommand{\opn}{\stackrel{\mathfrak{n}}{\cdot}}
\newcommand{\opx}{\stackrel{x}{\cdot}}

\newcommand{\tr}{\ \mbox{tr}}

\newcommand{\Ad}{\ \mbox{Ad}}

\newcommand{\ad}{\ \mbox{ad}}

\renewcommand{\contentsname}{ }

\title{Chemical concrete machine}

\author{Marius Buliga \\ 
\\
Institute of Mathematics, Romanian Academy \\
P.O. BOX 1-764, RO 014700\\
Bucure\c sti, Romania\\
{\footnotesize Marius.Buliga@imar.ro}}

\date{This version: 26.09.2013}

\maketitle

\begin{abstract}
The chemical concrete machine is a graph rewriting system which uses only local moves (rewrites), seen as chemical reactions involving molecules which are graphs made up by 4 trivalent nodes.  It is Turing complete, therefore it might be used as a model of computation in algorithmic chemistry. 
\end{abstract}

\section{Introduction}

Moe-Behrens explains \cite{moeb} that  there is a trend to use the architecture of a silicon computer for building a biological computer.  Whatever happens in a living cell which resembles to computation (in a more vague sense than Turing computation) is not necessarily implemented as in a silicon computer, though. Instead, synthetic biology tries to import concepts from silicon computers into the bio realm, mainly because these concepts are already familiar. Hence the accent on bits and boolean logic, although not exclusive (other studies concern for example membrane computing, or neural networks, or continuously evolving dynamical systems).  More important, all these studies seem to concentrate on imperative programming. 

Functional programming  is the other  main  programming paradigm, a very elegant one,  receiving more and more attention in the competition with the more well known imperative programming. Functional programming has its roots in the lambda calculus, one of the two pillars of computation, along with the Turing machine (of equivalent power, but with different philosophies behind). 

In 1992 Berry and Boudol \cite{bb} introduce the chemical abstract machine, which models asynchronous concurrent computations (in particular a concurrent lambda calculus) by using a chemical metaphor. The chemical abstract machine has states which are seen as chemical solutions  and the evolution from a state to another is ruled by chemical reactions, heating and cooling. Chemical solutions are  multisets of molecules, i.e. list of numbers of molecules of each type. 

Between 1994-1996, Fontana and Buss \cite{fonbas1} \cite{fonbas2} \cite{fonbas3}  introduce the idea that lambda calculus is a kind of natural formalization of the bare bones of  chemistry.  Individual molecules are seen as lambda terms, reactions between  molecules are seen as the application operation in lambda calculus, and  the abstraction operation from lambda calculus "captures the role of functional groups embedded in the reactants", \cite{fonbas2}  p. 11.  Fontana and Buss use lambda calculus with eta reduction, i.e. the terms (or molecules) are identified with functions, in the sense that the molecule $ A$ is identified with the function $ B \mapsto AB$. 

The work of Fontana and Buss started the "algorithmic chemistry" research field. In a sense, this can be seen as a part of an older research subject, the very difficult one, of course, concerning the organization and functioning of living systems. They cite  in \cite{fonbas2} the seminal paper  by Varela, Maturana and Uribe \cite{varela}.

The chemical concrete machine  is a graphic formalism which uses a chemical metaphor. It is a modification of graphic lambda calculus, introduced in \cite{bgraph}.  (See also the graphic lambda calculus  \href{http://chorasimilarity.wordpress.com/graphic-lambda-calculus/}{web tutorial}. Moreover, there is already a \href{http://chorasimilarity.wordpress.com/chemical-concrete-machine/}{web tutorial for the chemical concrete machine}.) 

The graphic lambda calculus has been constructed for reasons independent from any of the subjects evoked until now. It is a graph rewriting system which has a sector which is equivalent with untyped lambda calculus, but it  has other interesting sectors as well, like the one which is equivalent with knot diagrams, or the one which contains finite differential calculus in spaces with dilations \cite{buligadil1}, i.e. emergent algebras \cite{buligairq}. It is therefore Turing complete, but in order to achieve this, it uses a GLOBAL FAN-OUT move. Moves (i.e. graph rewrites) can be characterized as local, when they involve a fixed number of nodes and arrows in the graphs, or global, otherwise. 

By modifying the graphic lambda calculus, we obtain the chemical concrete machine, which is Turing complete by using only local moves. Another difference from graphic lambda calculus is that we reformulate the moves as chemical reactions between molecules (represented by trivalent graphs made by 4 types of trivalent nodes). Thus a computation in the chemical concrete machine is simply a chemical reaction network involving molecules and chemical reactions from the formalism. 

There are also some differences between the algorithmic chemistry of Fontana and Buss and the chemical concrete machine. The most important one is that  that  the application and the abstraction operations from lambda calculus are, in the chemical  concrete machine, trivalent nodes (say, like atoms which compose the molecules), and moves (corresponding to reductions in lambda calculus) are assimilated with enzymes, thus reductions are certain chemical reactions. Also, the chemical concrete machine can be used for implementing lambda calculus without extensionality, mainly because the eta reduction appears as a global move in graphic lambda calculus, which does not seem possible to be replaced by local moves, as it has been done with the GLOBAL FAN-OUT.

 This leads  to the last point: it is intriguing to search if "extreme" functional programming (i.e. functional programming without extensionality) is easier to be implemented in biological computing than imperative programming.  Already, lambda calculus looks like a kind of chemistry. In the chemical concrete machine the somehow esoteric application and abstraction are made concrete. Moreover,  because the chemical concrete machine uses the graphic lambda calculus formalism, it follows  that simple spatial manipulations, like grouping  molecules together, releasing them in some order or all of them simultaneously,  are more straightforward to achieve  in graphic lambda calculus than in the usual lambda calculus.

Another possibility would be to try to implement the chemical concrete machine in silico, which will be discussed in another article. 

\paragraph{Acknowledgments.} I benefited from various discussions with the following (in no particular order): Stephen P. King, Eugenio Battaglia, Gerd Moe-Behrens, Louis Kauffman, ine (nickname in comments made at the chorasimilarity blog), Lucius Meredith, Mike Stay.

\section{Chemical concrete machine}

The chemical concrete machine  is a graph rewriting system, derived from graphic lambda calculus \cite{bgraph}, which can be seen as a model of chemical, or why not biological, computing.

The distinction from other such models is that it works with very specific graphs, called "molecules", which interact in very specific ways. This is the source of  the word "concrete" in the denomination, as opposed to the chemical abstract machine of  Berry and Boudol \cite{bb}.

I hope it can be implemented in reality (that is in a real chemical reaction network), but nevertheless it can be seen as a proof of principle for the fact that the chemistry of a very small number of molecules can manifest Turing completeness. In this respect, the chemical concrete machine appears as  applied algorithmic chemistry  in a virtual world with chemistry rules which are made up by a mathematician.

{\bf Molecules. }    In the following I shall use the name "molecule" in the following sense. A molecule is seen as made by other smaller molecules (for example atoms are molecules) which are connected by chemical bonds, or by other molecules called "arrows". Therefore, a molecule is seen as a graph with nodes which are smaller molecules and arrows which connect those nodes. A collection of molecules is a molecule.  

The building block of molecules are the following elementary, or essential ones. 

\vspace{.5cm}

\centerline{\includegraphics[width=  110mm]{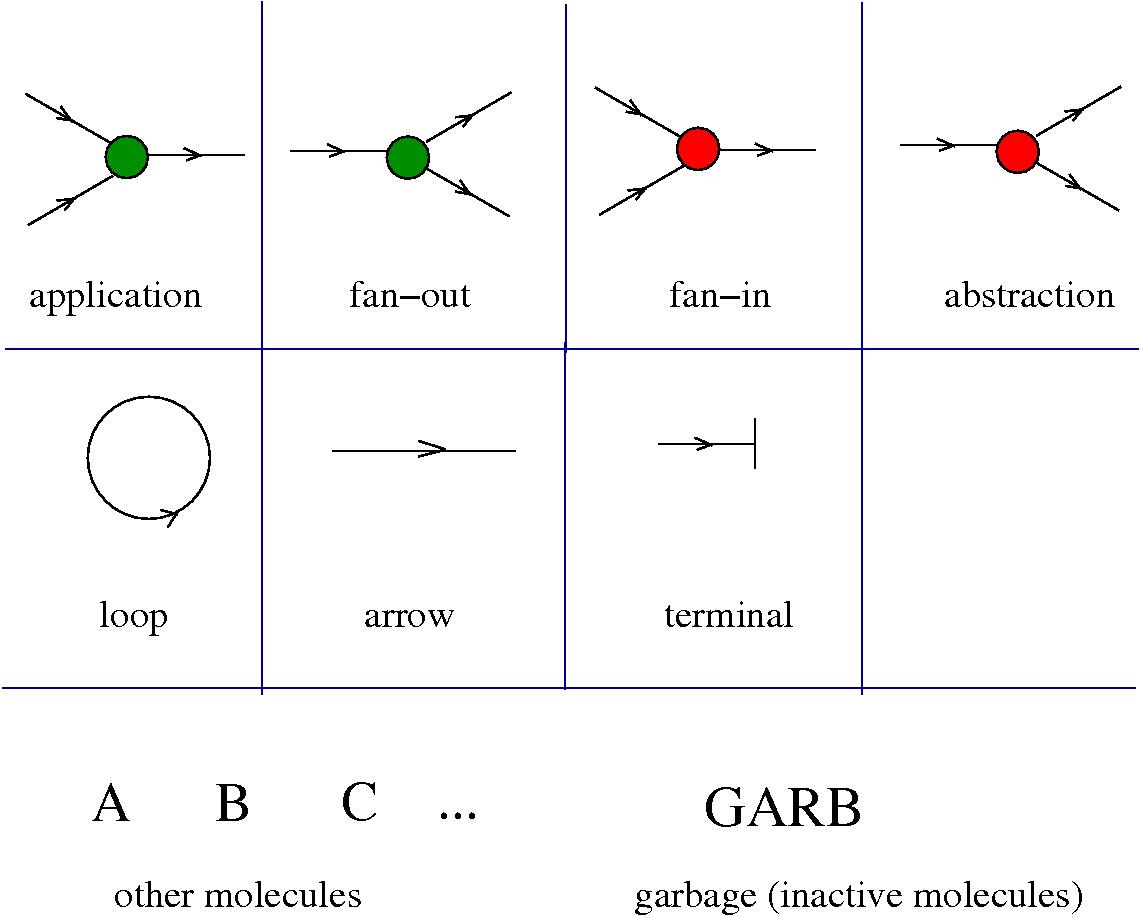}}

\vspace{.5cm}

On the first row you see the list of essential molecules, named respectively
\begin{enumerate}
	\item[-]"application", corresponding to the application gate in graphic lambda calculus, or to the application operation in lambda calculus
	\item[-]"fan-out"  ,  corresponding to the FAN-OUT gate in graphic lambda calculus  
	\item[-]"fan-in", which will replace the $\varepsilon$ gate from graphic lambda calculus  
	\item[-] "abstraction", corresponding to the $\lambda$ gate in graphic lambda calculus, or to the lambda abstraction from lambda calculus. 
\end{enumerate}

On the second row we see a loop, an arrow (which are considered molecules, as written before), and a molecule called "terminal", which corresponds in graphic lambda calculus to the termination gate.

We shall work also with a list of "other molecules", which can react (or not) one with another, this is left to the choice of the user of the chemical concrete machine, which can add more reactions in the formalism. There is also a collection $GARB$ of garbage.

\begin{definition} The set $MOLECULES$ of molecules is formed by all locally planar graphs which can be formed by using the essential molecules (and the "other molecules" with unspecified valences) as nodes, and arrows obtained by connecting the essential molecules such that they respect the arrows orientation. The graphs can have free arrows (with one of the ends of an arrow not connected to any other molecule), also they can have several connected components (i.e. a finite union of molecules is a molecule). Loops and arrows are molecules. 
\label{defmol}
\end{definition}

The rules of building molecules are exactly the ones for building graphs in the set $ GRAPH$ described in section 2 \cite{bgraph}, or go directly to the \href{http://chorasimilarity.wordpress.com/graphic-lambda-calculus/}{graphic lambda calculus tutorial}, with the understanding that for the chemical concrete machine we use colored nodes instead of the gates from graphic lambda calculus, as described in the following figure. The only new thing is that we admit also a list $ A, B, C, ...$ of unspecified nodes with unspecified valences, which model "other molecules". 

\vspace{.5cm}

\centerline{\includegraphics[width=  80mm]{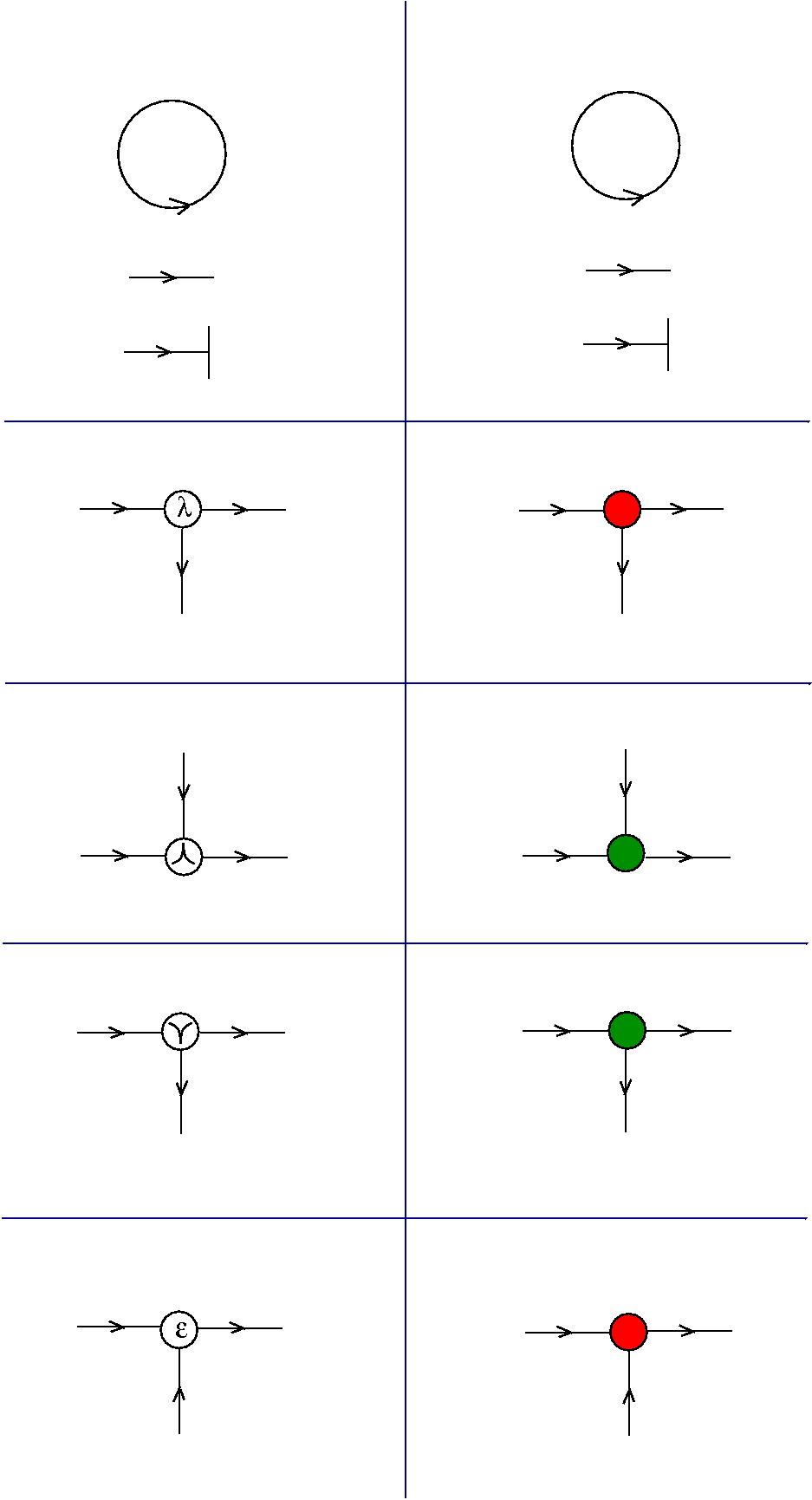}}

\vspace{.5cm}

Here are some examples of  molecules.

\vspace{.5cm}

\centerline{\includegraphics[width=  90mm]{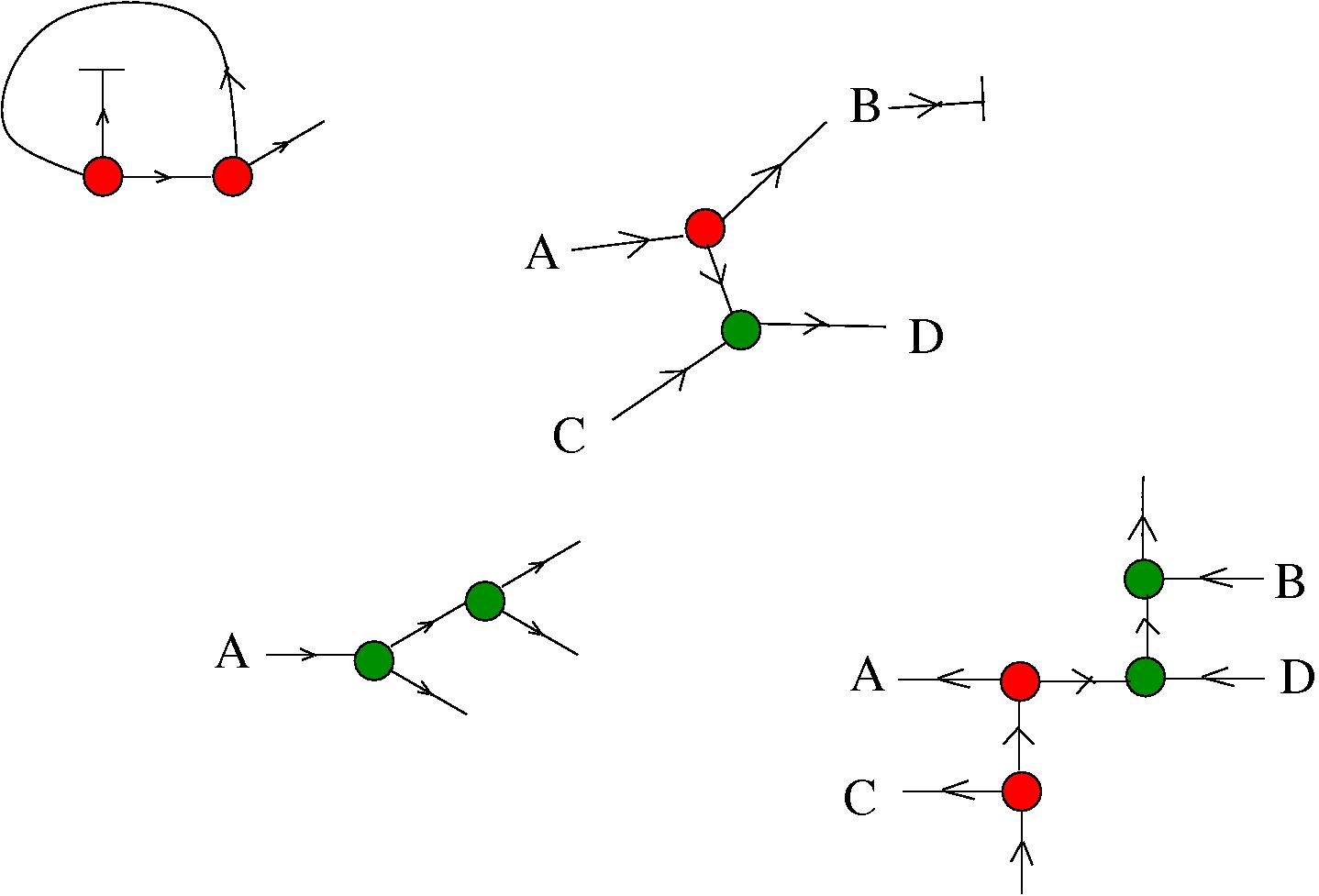}}

\vspace{.5cm}

We imagine that all molecules float inside a container. There may be several copies of the same molecule in the container.

{\bf Enzymes and moves.} Besides molecules, I shall also use "enzymes". A chemical reaction will always involve  a molecule and an enzyme.  Enzymes are not molecules. Instead, enzymes are names of moves (graph rewrites).

\vspace{.5cm}

\centerline{\includegraphics[width=  90mm]{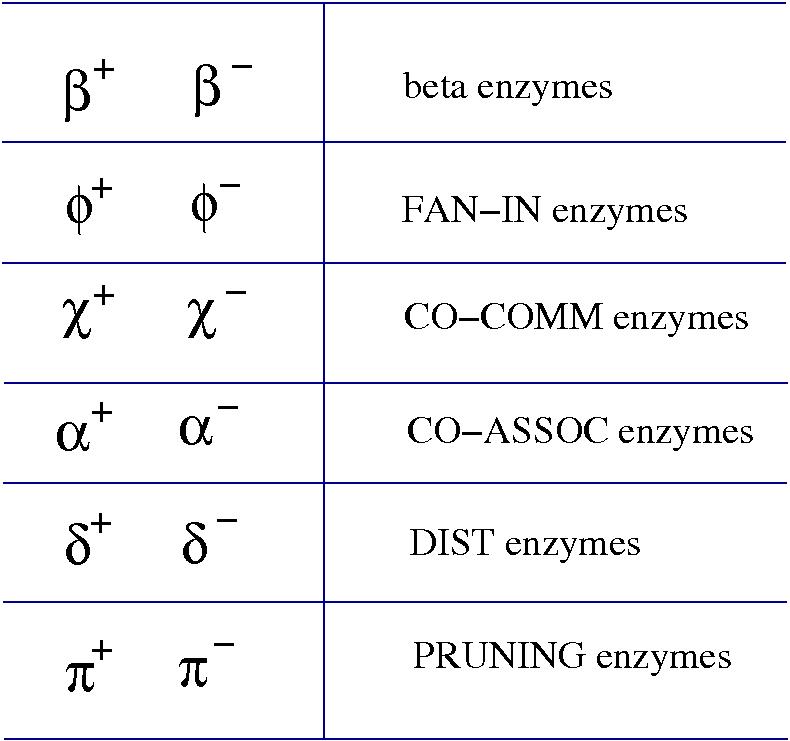}}

\vspace{.5cm}

What is to be noticed is that any enzyme comes in two flavors: "+" and "-". The reason for this is the following. Any chemical reaction which involves a molecule and an enzyme will correspond to a move in the list of moves of the chemical concrete machine formalism. The list of moves of the chemical concrete machine, which will be described soon, contains the local moves of graphic lambda calculus which involves the fan-out, application and abstraction gates, supplemented with a list of moves related to the fan-in gate, which replace the emergent algebra moves from the graphic lambda calculus. All the moves of the graphic lambda calculus are bi-directional.  Therefore, any such  move can be seen as a pair of unidirectional moves. By convention, the moves from left to right are denoted by names of enzymes with "+" and the moves from right to left are denoted by names of enzymes with "-". 

We cannot simply write chemical reactions as 

\centerline{{\large \em molecules + enzyme = molecules + GARB}} 

because these molecules may be complicated beasts (graphs) and because, as we shall see, the enzymes prefer to react with certain patterns (subgraphs) of molecules. For specifying how the reaction takes place we need:
\begin{enumerate}
	\item[-]molecules  
	\item[-]an enzyme  
	\item[-]and a "reaction site", which is a small part of the initial collection of  molecules.  
\end{enumerate}
The reaction site have to be present in the molecule, otherwise the reaction cannot happen. 

In order to explain how to translate from the language of  chemical reactions and enzymes to the language of moves of the chemical concrete machine, let's contemplate  the following figure.  I consider two examples of reactions (which will turn out to correspond to \href{http://chorasimilarity.wordpress.com/2012/12/13/the-graphic-beta-move-with-details/}{graphic beta moves}  in the realm of graphic lambda calculus).

\centerline{\includegraphics[width= 130mm]{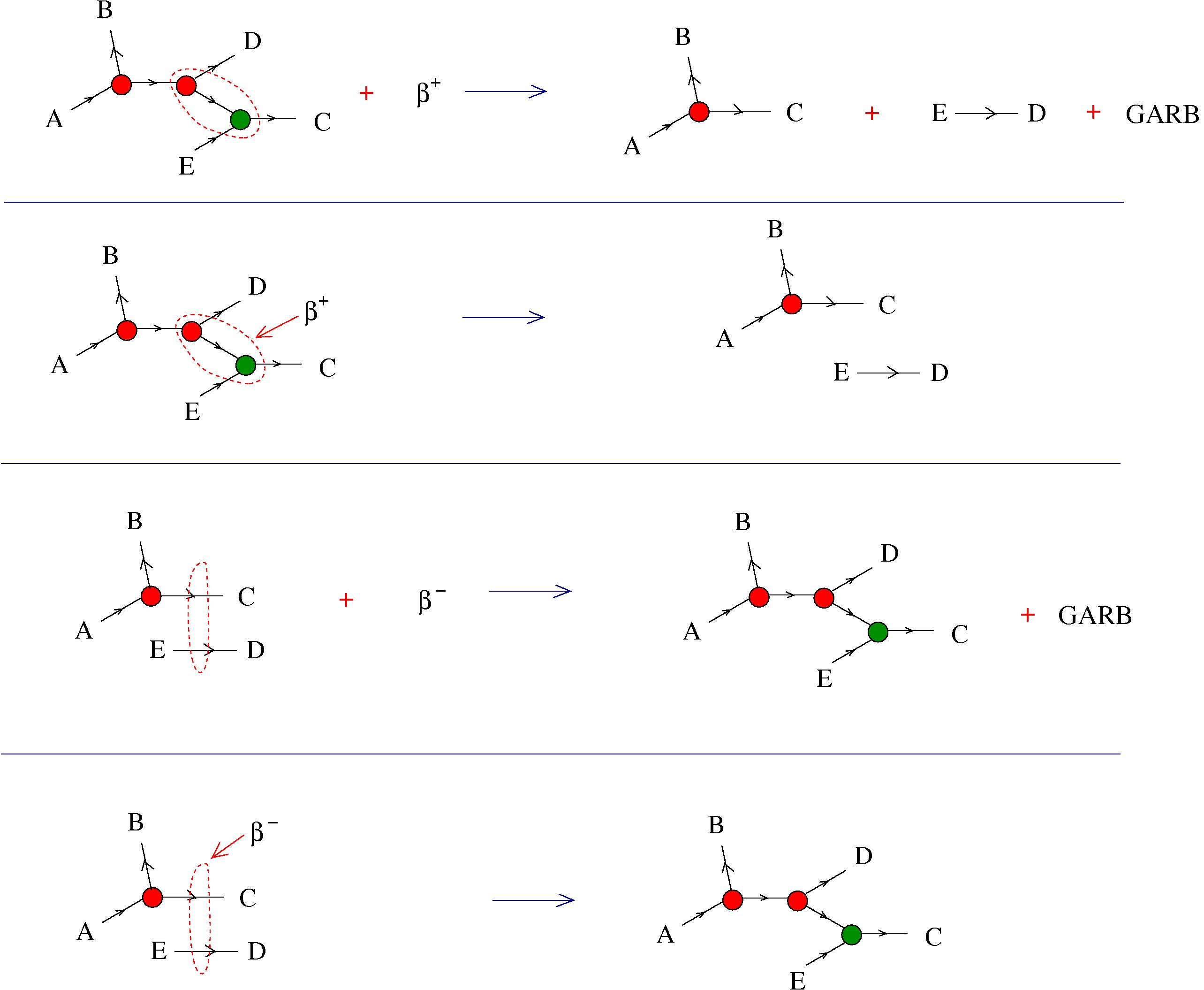}} 

In the first row we see a reaction between a molecule and the enzyme $ \beta^{+}$, which results into two other molecules and some GARB.  There is a small region in the initial molecule, marked by a dashed red closed curve, which represents a reaction site for the $ \beta^{+}$ molecule.

In the second row is written the same reaction, but in a simpler form. The red "+" sign is eliminated, the two molecules which are obtained are juxtaposes, as if they are floating in the 3D container, and the GARB is ignored. Moreover, the enzyme $ \beta^{+}$ points towards the reaction site.

The rows 3 and 4 describe another reaction. At closer inspection, it's a reaction which can be interpreted as the inverse of the first one.  Let's examine directly the 4th row (which is obtained from the 3rd row  by the same procedure as the 2nd row was obtained from the 1st).  The reaction site of the enzyme $ \beta^{-}$ is a pair of arrows from two different molecules, The resulting molecule is the same as the initial molecule from the previous reaction.

\begin{definition} Here is the list of moves, with the names taken from graphic lambda calculus (where it is the case), by using the dictionary for translation (links to the web tutorials are given). 
\begin{enumerate}
	\item[-]\href{http://chorasimilarity.wordpress.com/2012/12/13/the-graphic-beta-move-with-details/}{graphic beta move}  and \href{http://chorasimilarity.wordpress.com/2013/07/13/local-fan-in-eliminates-global-fan-out-i/}{FAN-IN move} :  
\end{enumerate}

\vspace{.5cm}

\centerline{\includegraphics[width=  110mm]{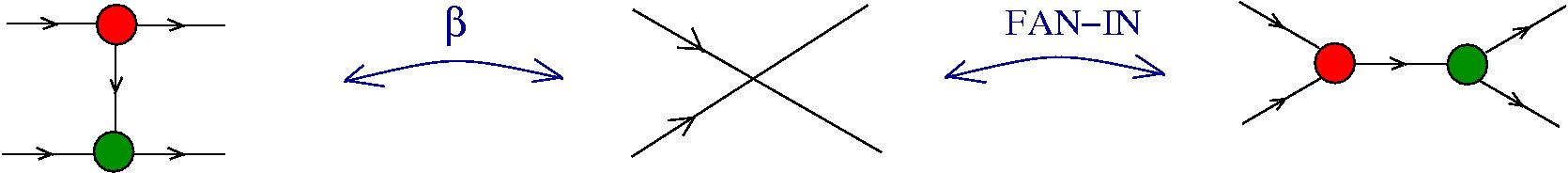}} 

\vspace{.5cm}

\begin{enumerate}
	\item[-]local \href{http://chorasimilarity.wordpress.com/2012/12/17/fan-out-moves-co-comm-co-assoc-global-fan-out-local-fan-out/}{FAN-OUT moves} :  
\end{enumerate}

\vspace{.5cm}

\centerline{\includegraphics[width=  90mm]{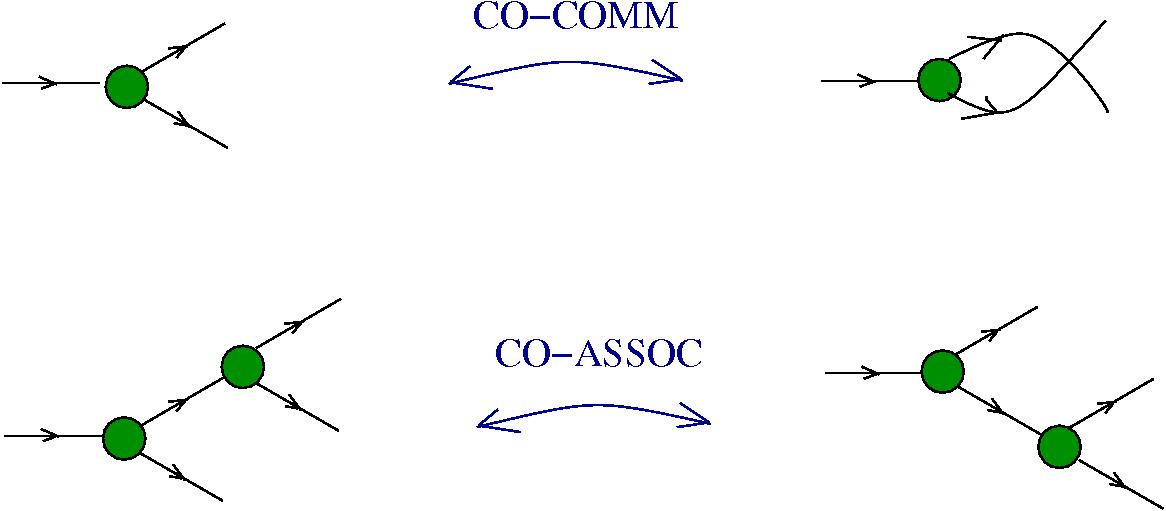}} 

\vspace{.5cm}

\begin{enumerate}
	\item[-]\href{http://chorasimilarity.wordpress.com/2013/07/13/local-fan-in-eliminates-global-fan-out-i/}{DIST}  moves ("dist" comes from the word "distributivity"):  
\end{enumerate}

\vspace{.5cm}

\centerline{\includegraphics[width=  90mm]{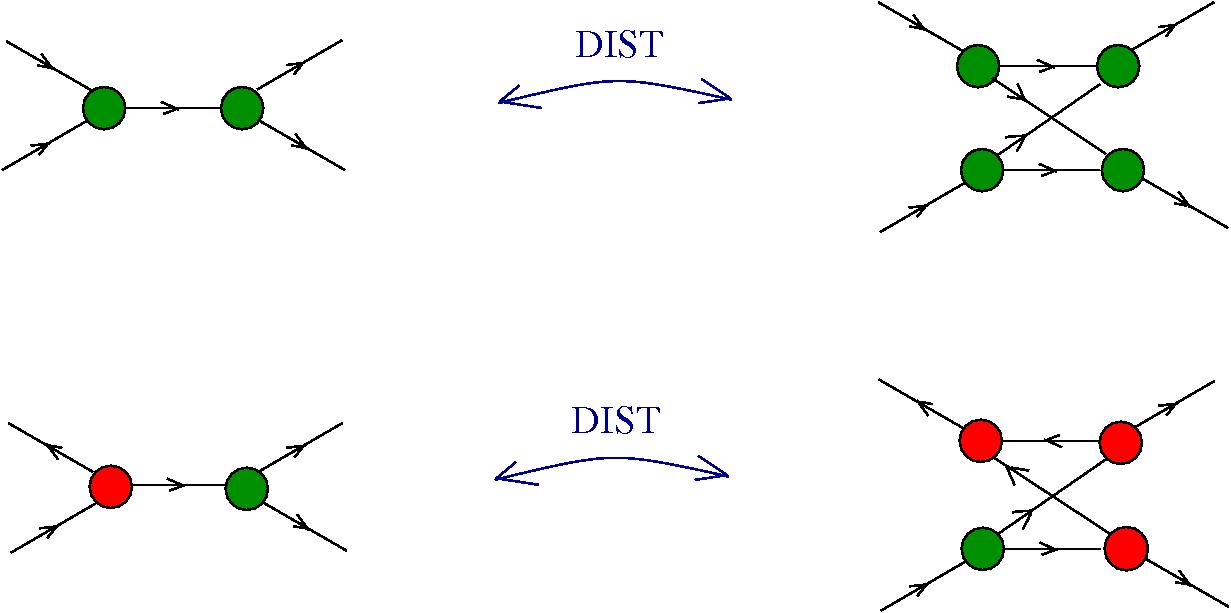}} 

\vspace{.5cm}

\begin{enumerate}
	\item[-]\href{http://chorasimilarity.wordpress.com/2012/12/17/pruning-moves/}{LOCAL PRUNING}  moves:  
\end{enumerate}

\vspace{.5cm}

\centerline{\includegraphics[width=  90mm]{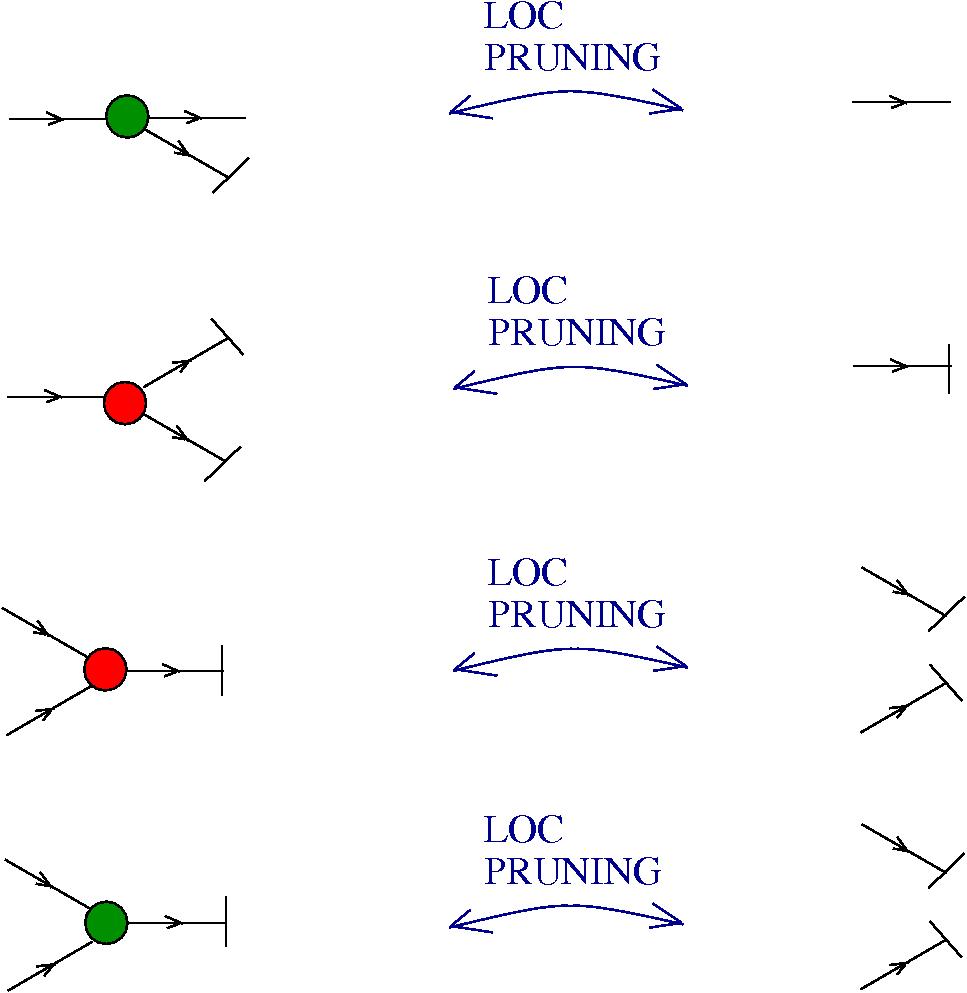}} 

\vspace{.5cm}

\begin{enumerate}
	\item[-]Elimination of loops (i.e. moving loops from and to $GARB$)
\end{enumerate}

\vspace{.5cm}

\centerline{\includegraphics[width=40mm]{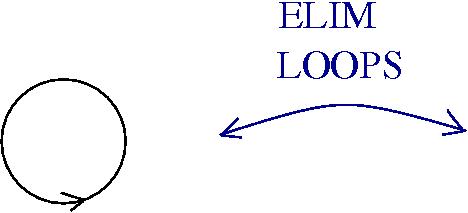}} 

\label{defmov}
\end{definition}

While all of these moves define the chemical concrete machine, some of them, seen in isolation, appear in other parts. I don't know if there is any, more profound, reason for this, but nevertheless here are some examples of such places: 
\begin{enumerate}
\item[-] in \cite{nima}, the merge move (2.28) section 2.6 looks like the pair of graphic beta move - fan-in move; the blow up move (13.1) section 13.1 looks even more like the graphic beta move, 
\item[-] in \cite{thurst}, the unzip move from section 3 looks almost alike the graphic beta move, 
\item[-] in \cite{kis}, after definition 3.1.3 of a bialgebra in a monoidal category, a DIST move and a LOC PRUNING move appear under the form of graphical identities. Of course, CO-COMM and CO-ASSOC moves can be seen as related to 
co-commutative and co-associative comonoids.  
\end{enumerate}

\section{Using the chemical concrete machine (I)}

Let's see what the chemical concrete machine can do.

\paragraph{Lists and locks.}  Suppose you have a family of molecules which you want to free in the medium in a given order. This corresponds to having a list of molecules, which is "read" sequentially. I shall model this with the help of the zipper from graphic lambda calculus.

Suppose that the molecules we want to manipulate have the form $ A \rightarrow A'$, with $ A$ and $ A'$ from the family of "other molecules" and $ \rightarrow$ an arrow.  Here are three zippers. 

\vspace{.5cm}

\centerline{\includegraphics[width=  90mm]{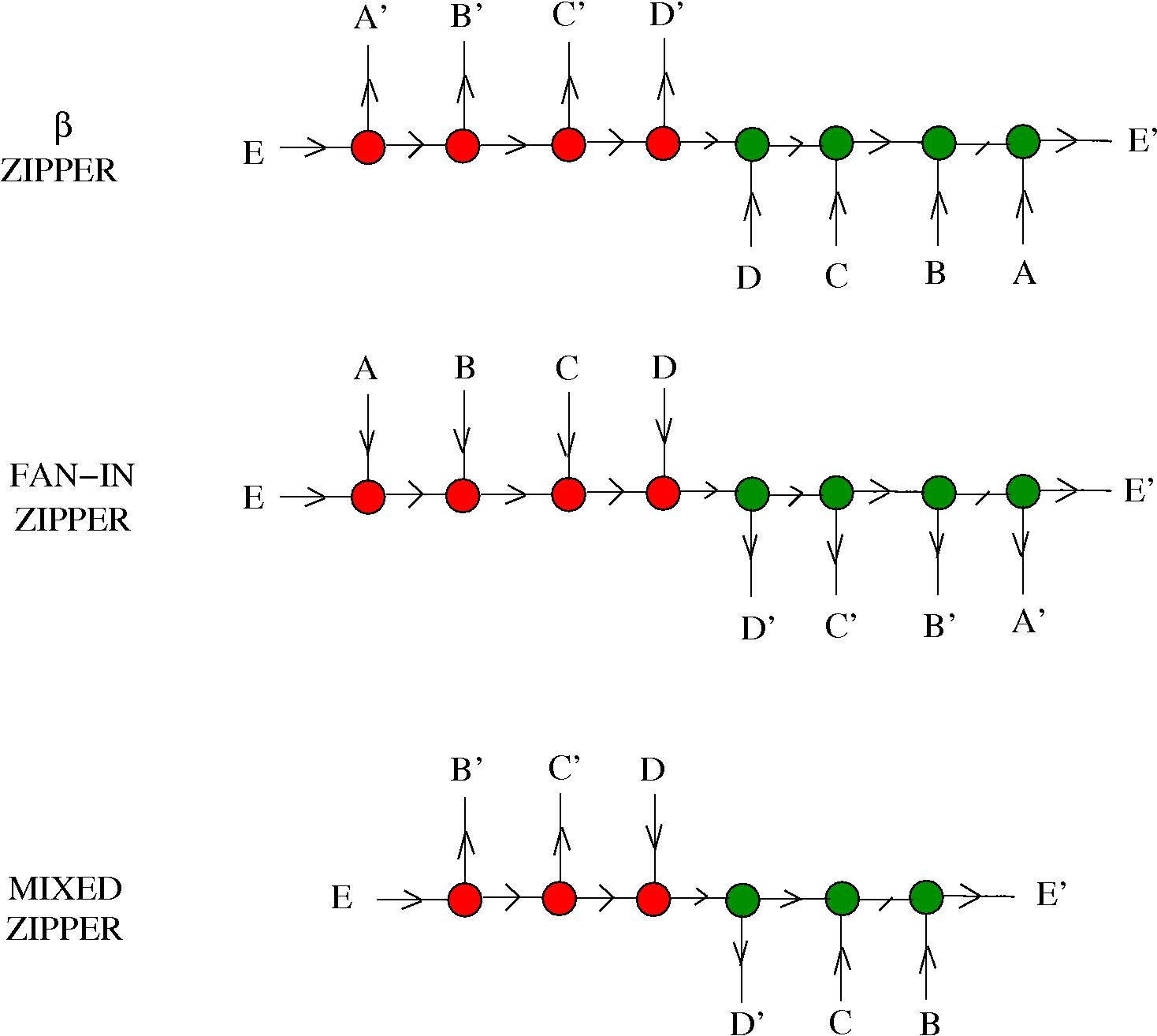}} 

\vspace{.5cm}

The first zipper, called a $ \beta$ zipper, behaves in the following way. In the presence of $ \beta^{+}$ enzymes, there is only one reaction site available, namely the one involving the red and green nodes in the neighbourhood of the $ D, D'$. So there is only one reaction possible with a $ \beta^{+}$ enzyme, which has a a result the molecule $ D \rightarrow D'$ and a new, shorter $ \beta$ zipper. This new zipper has only one reaction site, this time involving nodes in the neighbourhood of $ C, C'$, so the reaction with the enzyme $ \beta^{+}$ gives $ C \rightarrow C'$ and a new, shorter zipper.  

\vspace{.5cm}

\centerline{\includegraphics[width=  90mm]{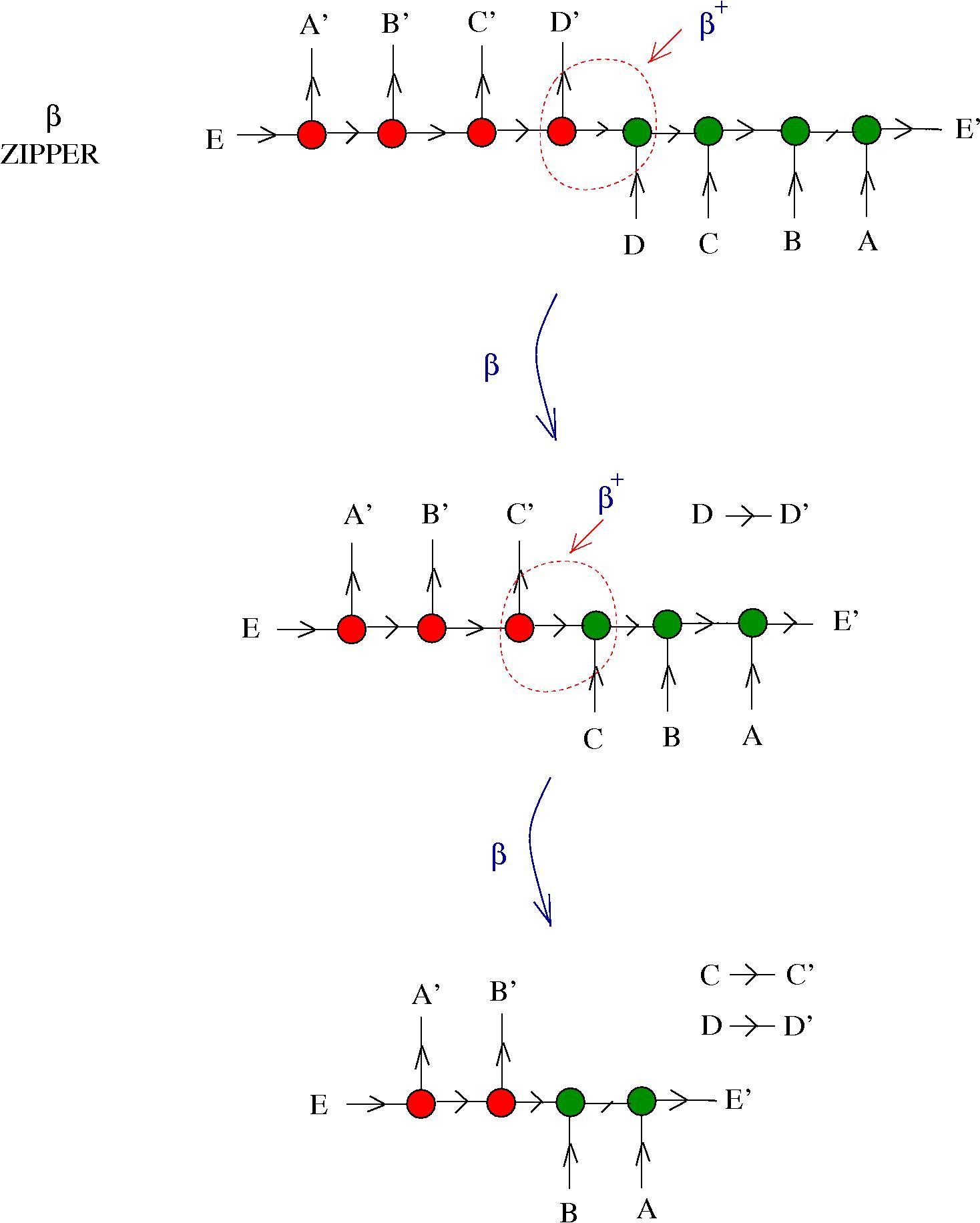}} 

\vspace{.5cm}

 The reaction continues like this, freeing in order the molecules $ B\rightarrow B'$, then $ A \rightarrow A'$ and $ E \rightarrow E'$.

The second zipper is called a FAN-IN zipper (or a $ \phi$ zipper). It behaves the same as the previous one, but this time in the presence of the FAN-IN enzyme $ \phi^{+}$.

In the case of a  mixed  zipper, he first molecule $ D \rightarrow D'$ is released only in the presence of a $ \phi^{+}$ enzyme, then we are left with a $ \beta$ zipper.

This can be used to lock zippers. Look for example at the following molecule:

\vspace{.5cm}

\centerline{\includegraphics[width=  90mm]{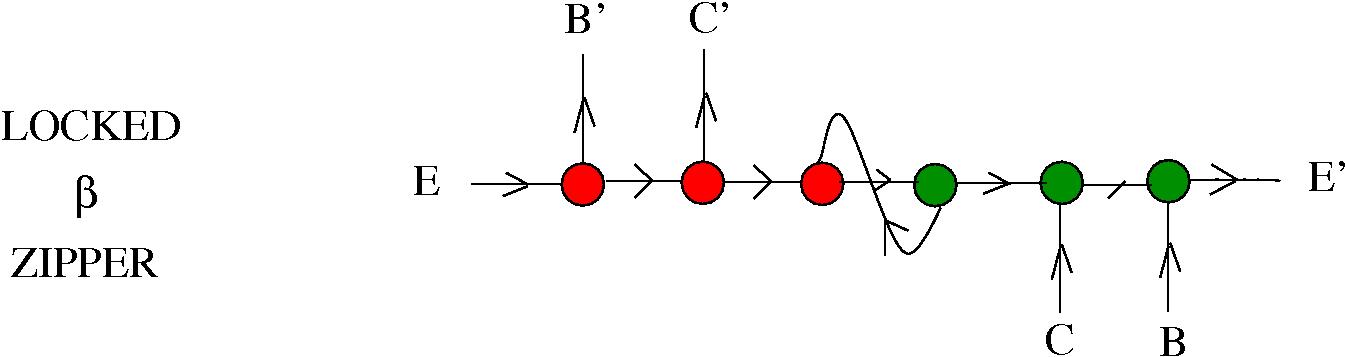}}

\vspace{.5cm}

called a locked $ \beta$ zipper. In the presence of only $ \beta^{+}$ enzymes, nothing happens, because there is no reaction site compatible with the $\beta^{+}$ enzyme. . If we add into the reactor also $ \phi^{+}$ enzymes, then the zipper unlocks, by releasing a loop (that's seen as garbage) and a $ \beta$ zipper which starts to react with $ \beta^{+}$ enzymes. 

\vspace{.5cm}

\centerline{\includegraphics[width=  90mm]{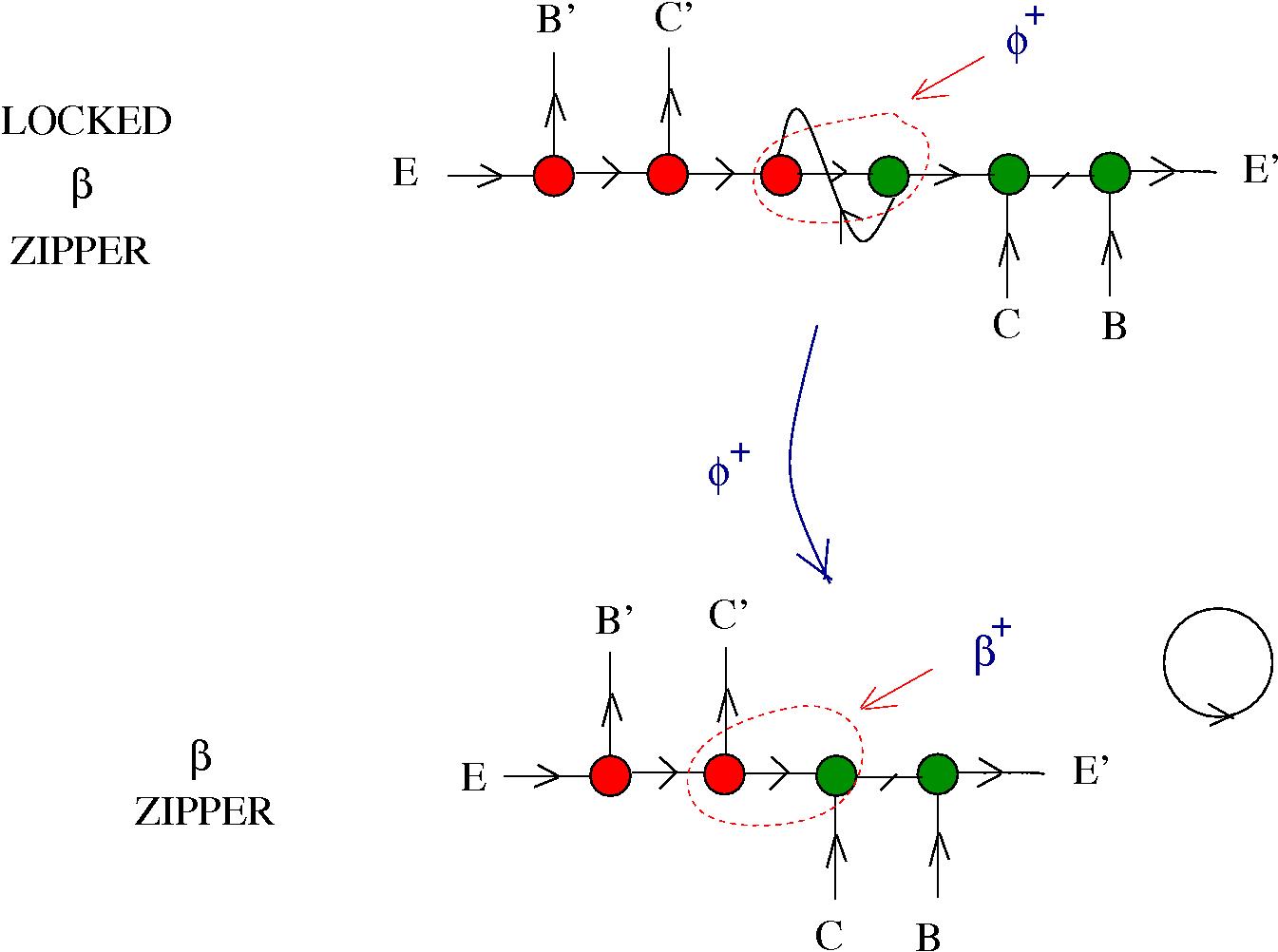}}

\vspace{.5cm}

The same idea can be used for keeping a molecule inactive unless both $ \phi^{+}$ and $ \beta^{+}$ enzymes are present in the reactor.  Say that w have a molecule $ A \rightarrow A'$ which is made inactive under the form presented in the following figure

\vspace{.5cm}

\centerline{\includegraphics[width=  100mm]{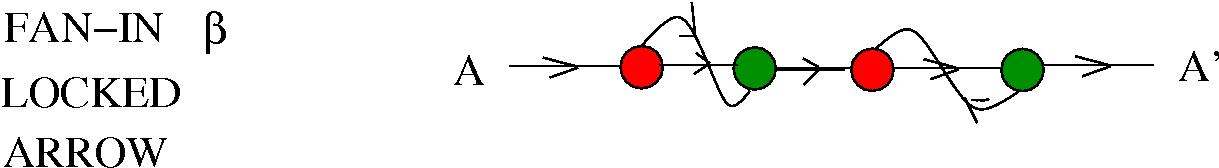}}

\vspace{.5cm}

The molecule is locked, but it has two reaction sites, one sensible to $ \beta^{+}$, the other sensible to $ \phi^{+}$. Both enzymes are needed for unlocking the molecule, but there is no preferred order of reaction with the enzymes (in particular these reactions can happen in parallel).

\vspace{.5cm}

\centerline{\includegraphics[width=  80mm]{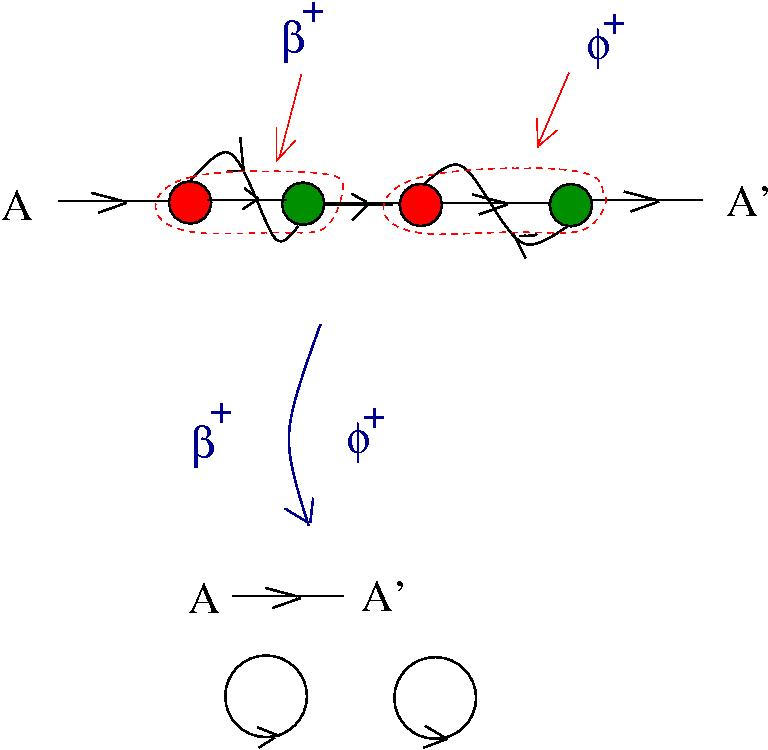}}

\vspace{.5cm}

\paragraph{Sets.} Suppose now that we don't want to release the molecules in a given order. We need to prepare a molecule which has several reaction sites available, so that multiple reactions can happen in parallel, as in the last example. Mathematically, that could be seen as a representation of the set of molecules we want to free, instead of the list of them.  This is easy, as described in the next figure:

\vspace{.5cm}

\centerline{\includegraphics[width=  110mm]{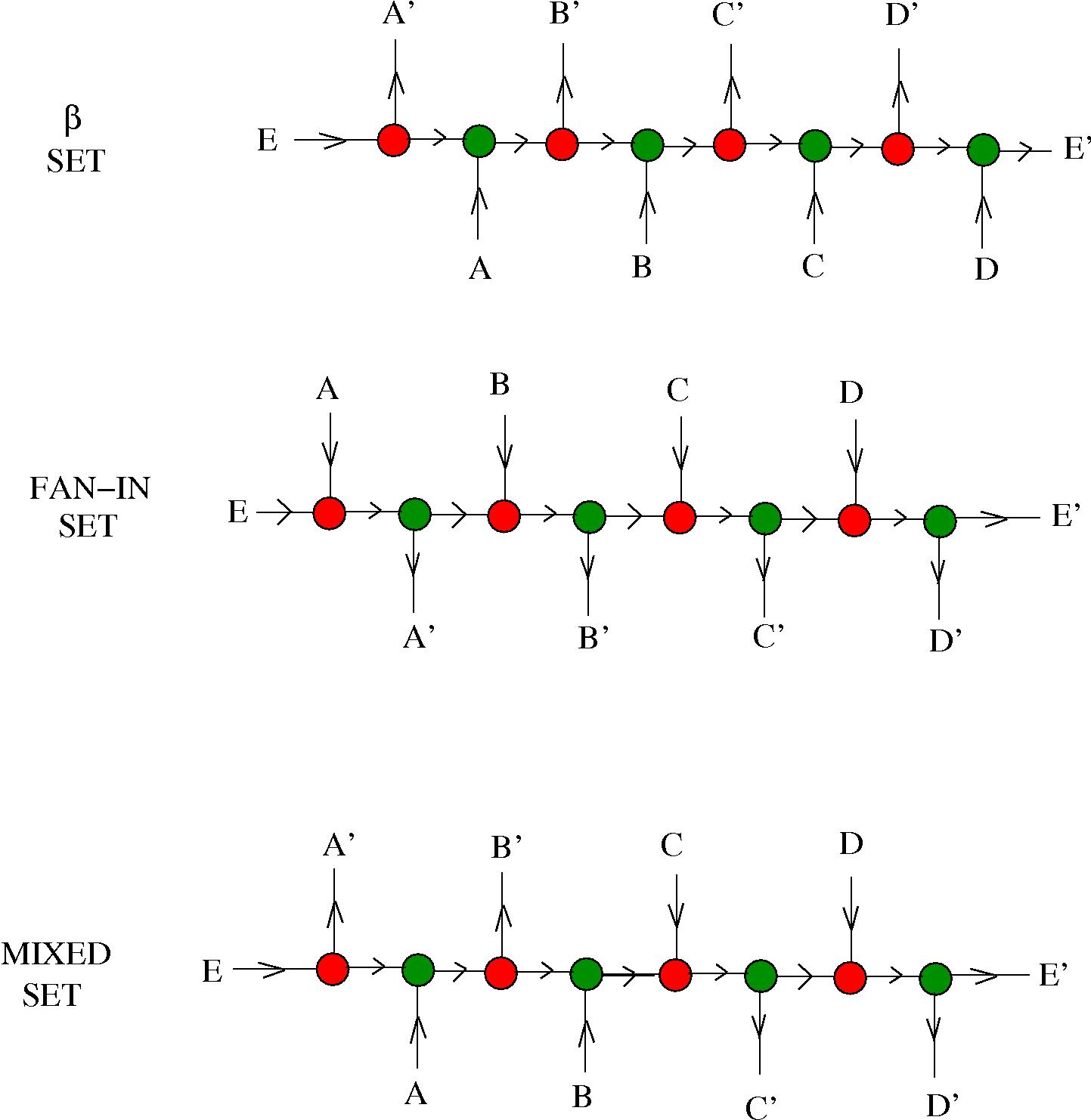}}

\vspace{.5cm}

On the first row we see what is called a $ \beta$ set. It has 4 possible reaction sites with the enzyme $ \beta^{+}$, therefore, in the presence of this enzyme, the molecules $ A \rightarrow A'$, ... , $E \rightarrow E'$ are released at the same moment. 

\vspace{.5cm}

\centerline{\includegraphics[width=  110mm]{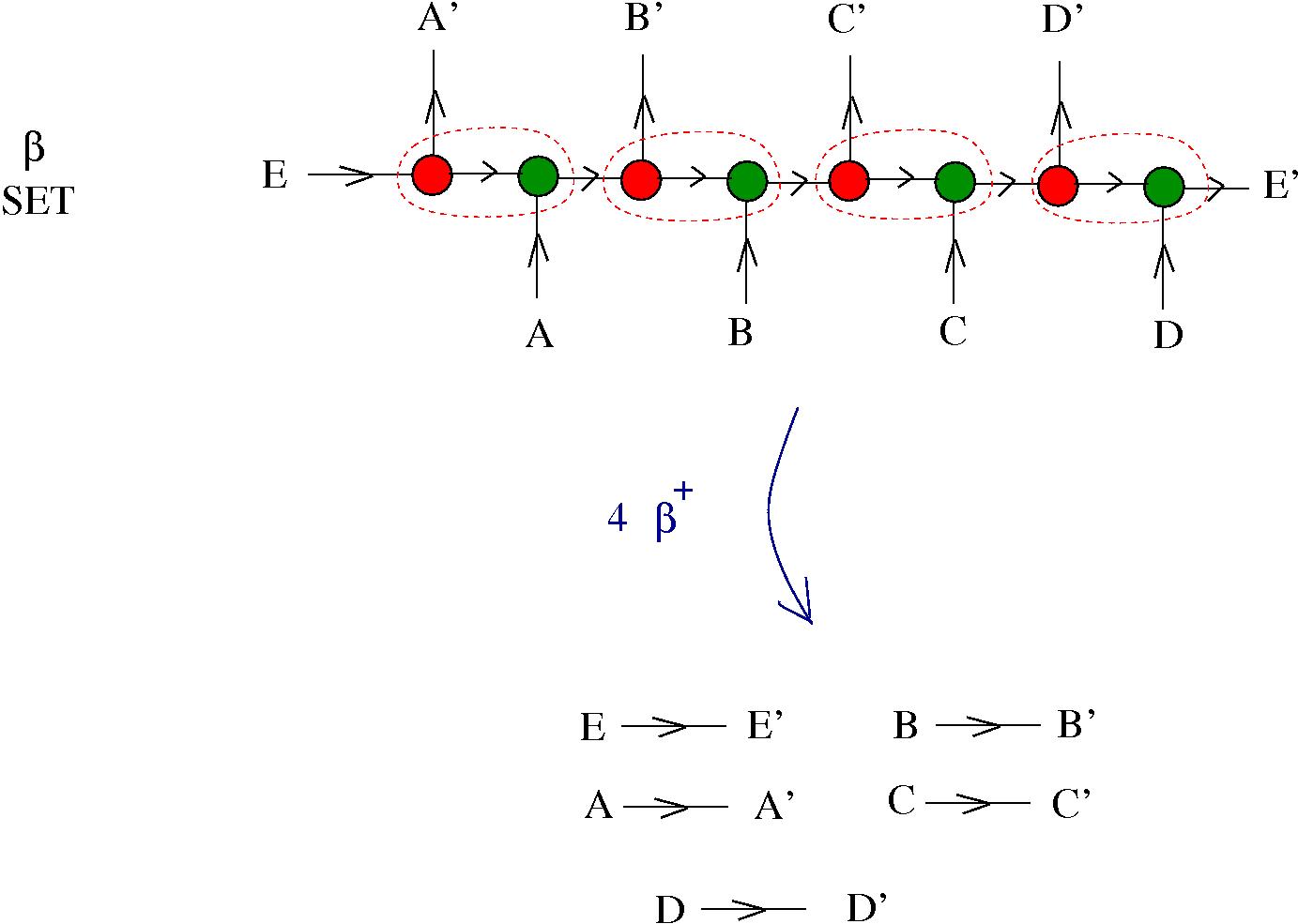}}

\vspace{.5cm}

A FAN-IN, or $ \phi$ set,  behaves the same as the previous one, but this time in the presence of the FAN-IN $ \phi^{+}$ enzyme.

Finally a mixed set releases different molecules, depending on the presence of $ \phi^{+}$ or $ \beta^{+}$ enzymes.

\paragraph{Pairs. }   As another example, here is a more involved molecule, which produces different pairs of molecules, according to the presence of $ \phi^{+}$ or $ \beta^{+}$ enzymes.

In the following figure we see how we model a pair of molecules, then two possible reactions a represented.

\vspace{.5cm}

\centerline{\includegraphics[width=  90mm]{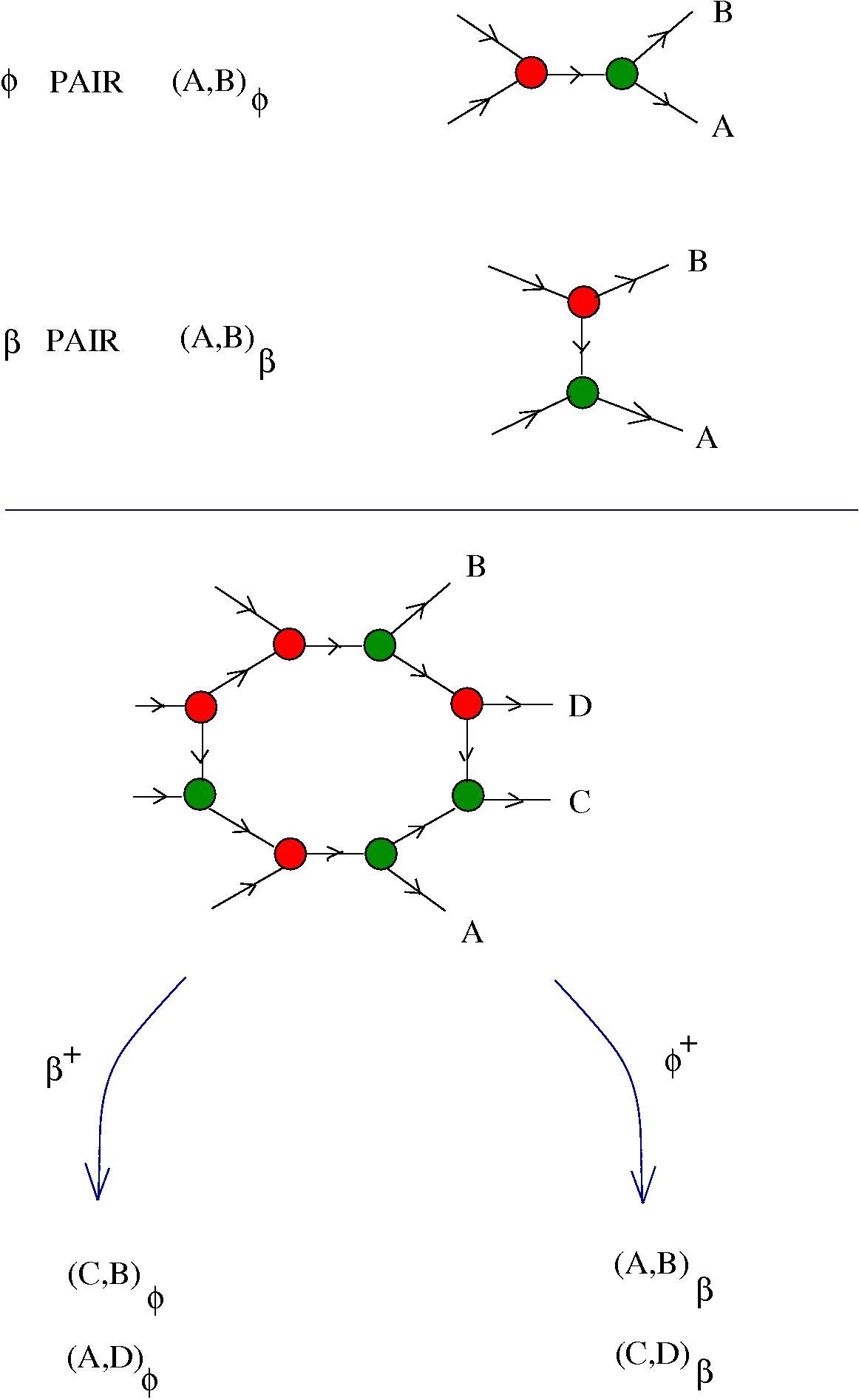}}

\vspace{.5cm}

The idea is that we can decide, by controlling the amount of $ \beta^{+}$ or $ \phi^{+}$, to couple $ A$ with $ D$ and $ C$ with $ D$, or to couple $ A$ with $ B$ and $ C$ with $ D$.

{\bf 4. Multipliers and comultipliers. }   Multipliers and co-multipliers  are molecules which self-multiply.  More precisely, in the next figure we see the definition of those:

\centerline{\includegraphics[width=  90mm]{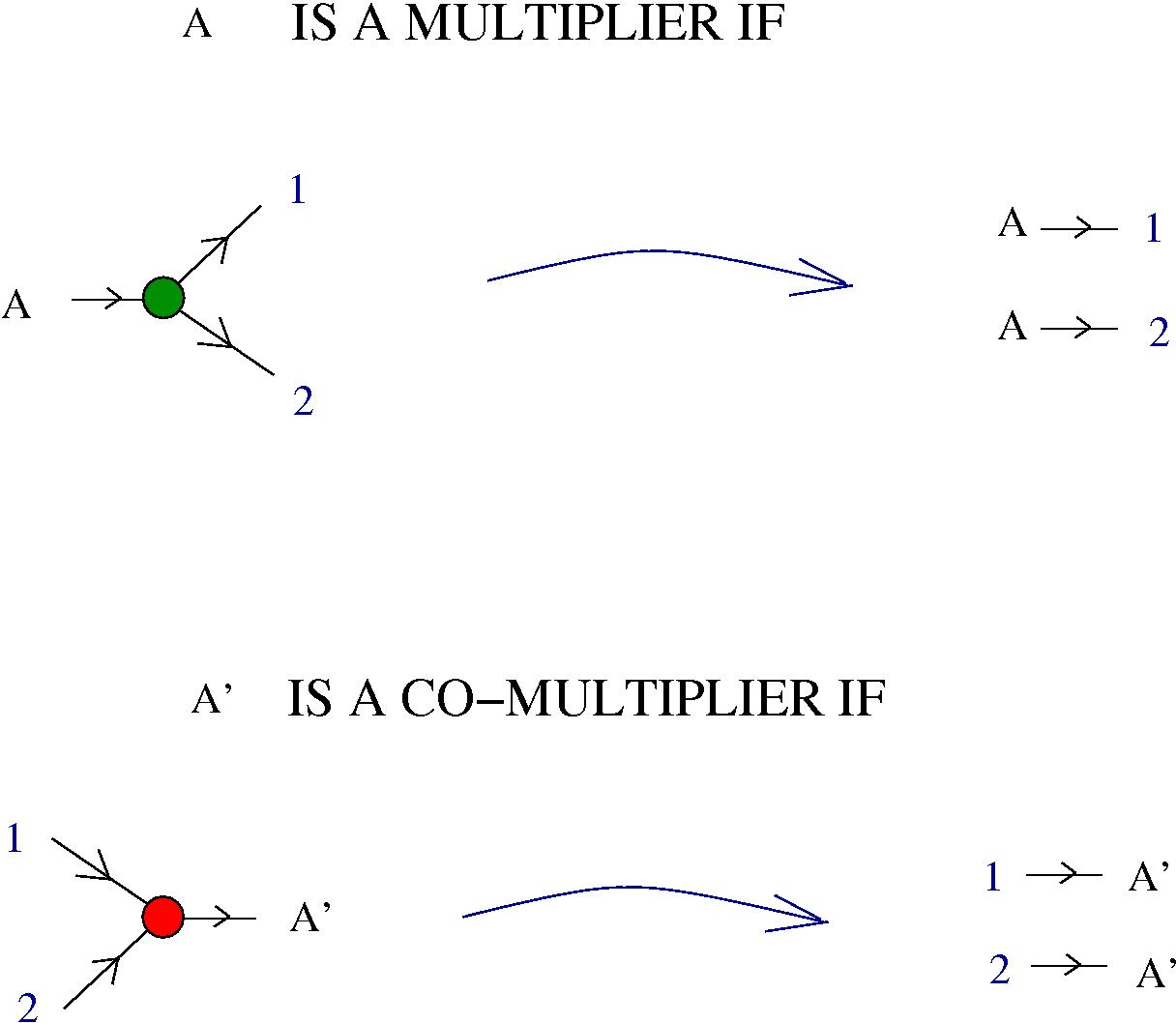}}

Here $ A$  and $ A'$ are molecules from the formalism of the chemical concrete machine and $ 1$ and $ 2$ are labels. The blue arrow means any finite sequence of chemical reactions (moves) from the formalism.

In certain configurations, zippers are multipliers. In the following figure we see what happens in the presence of DIST enzymes:

\centerline{\includegraphics[width=  90mm]{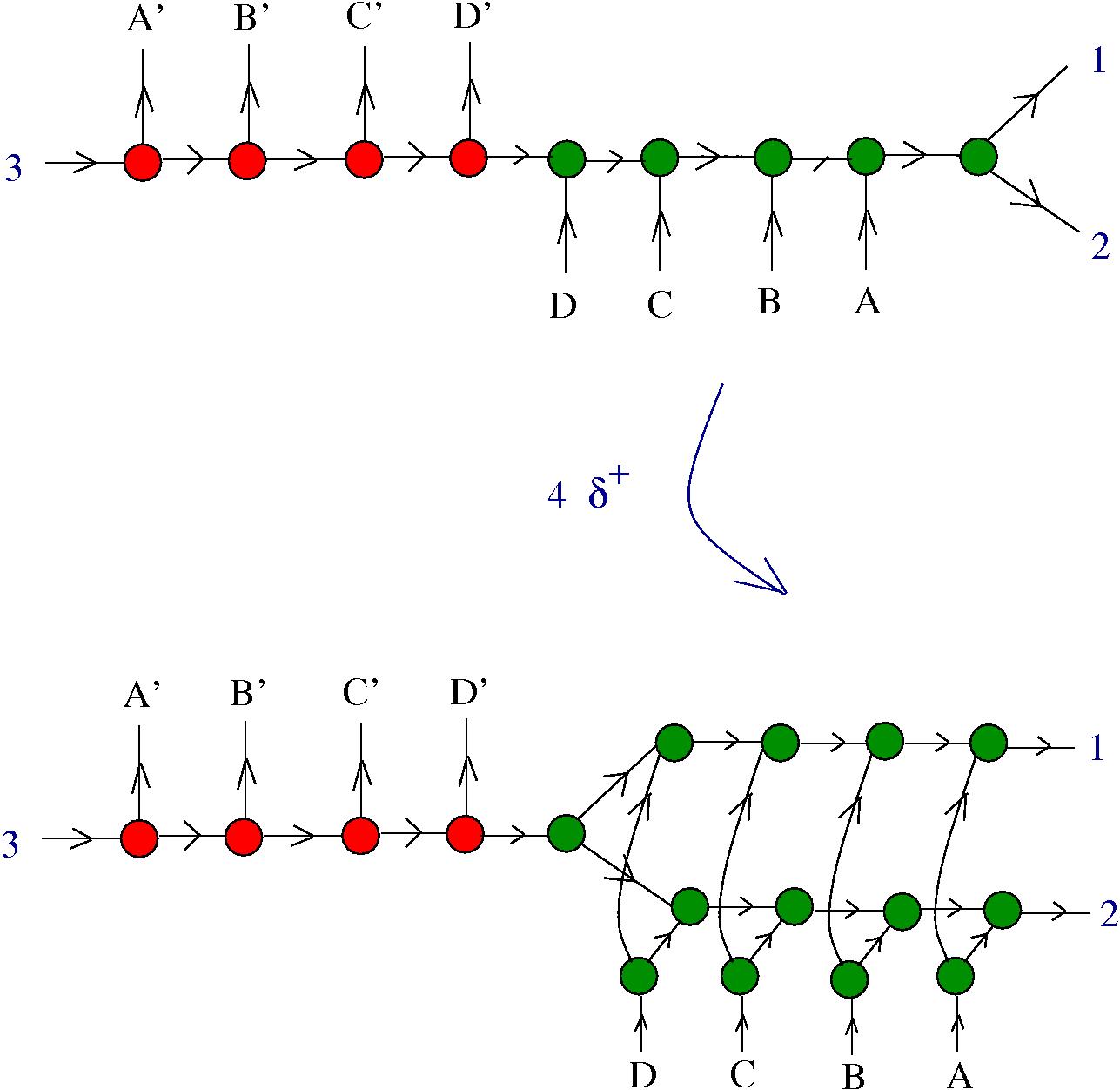}}

The reaction continues:

\centerline{\includegraphics[width=  90mm]{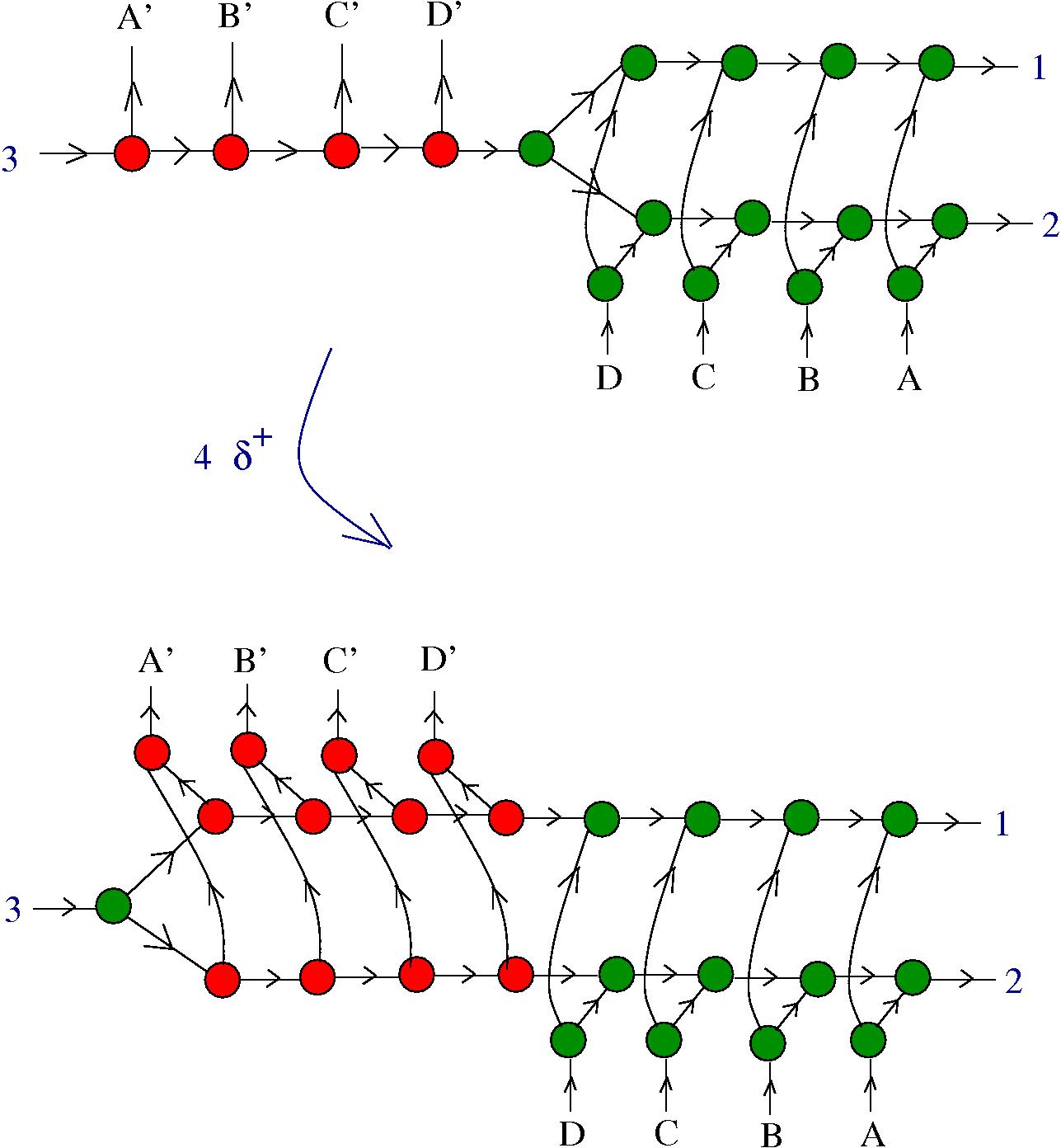}}

Now, the zipper multiplied into two zippers, but they are still connected.  We need more information about $ A, B, C, D$ and $ A', B', C', D'$.   Remark that:

\centerline{\includegraphics[width=  90mm]{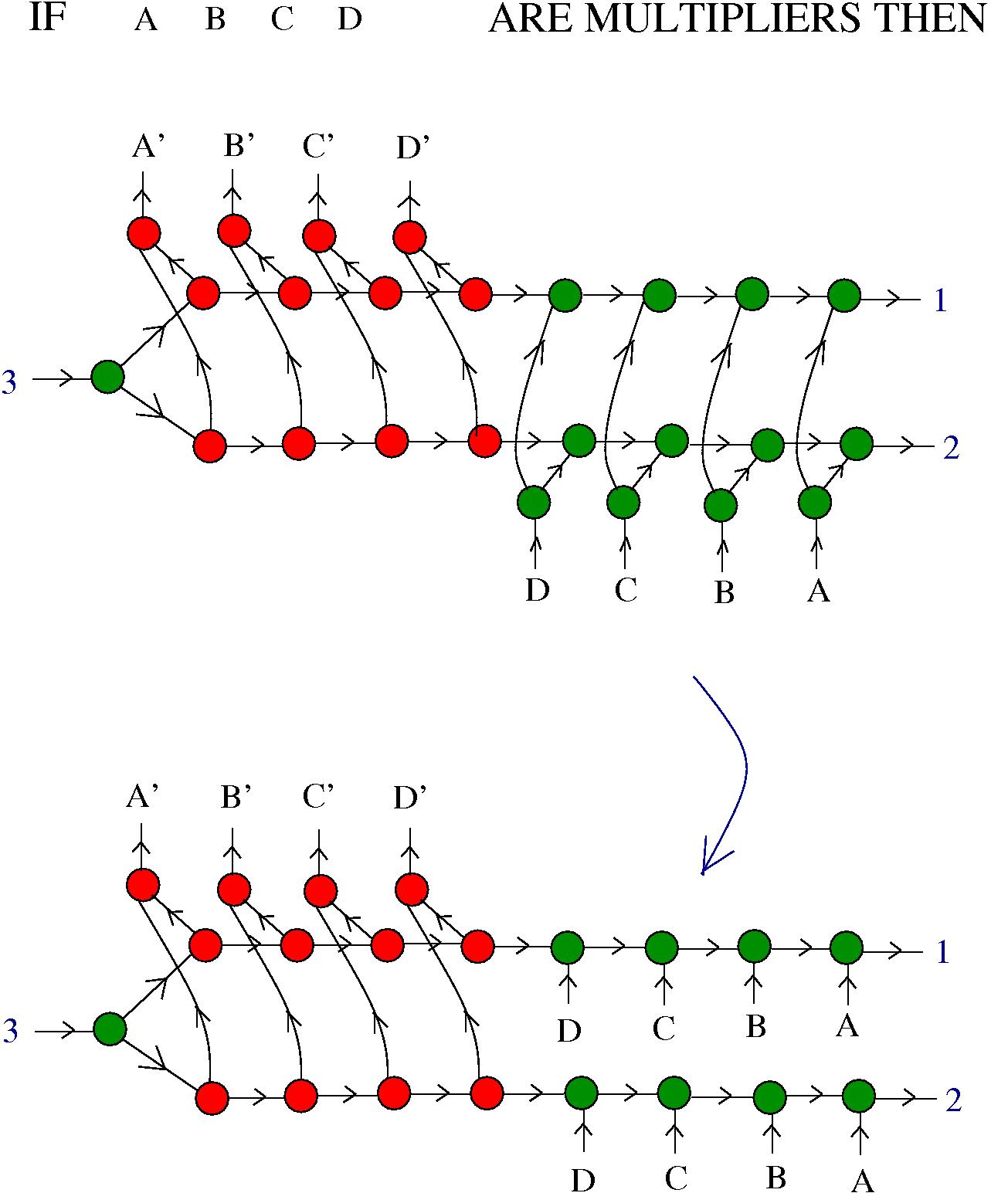}}

\centerline{\includegraphics[width=  90mm]{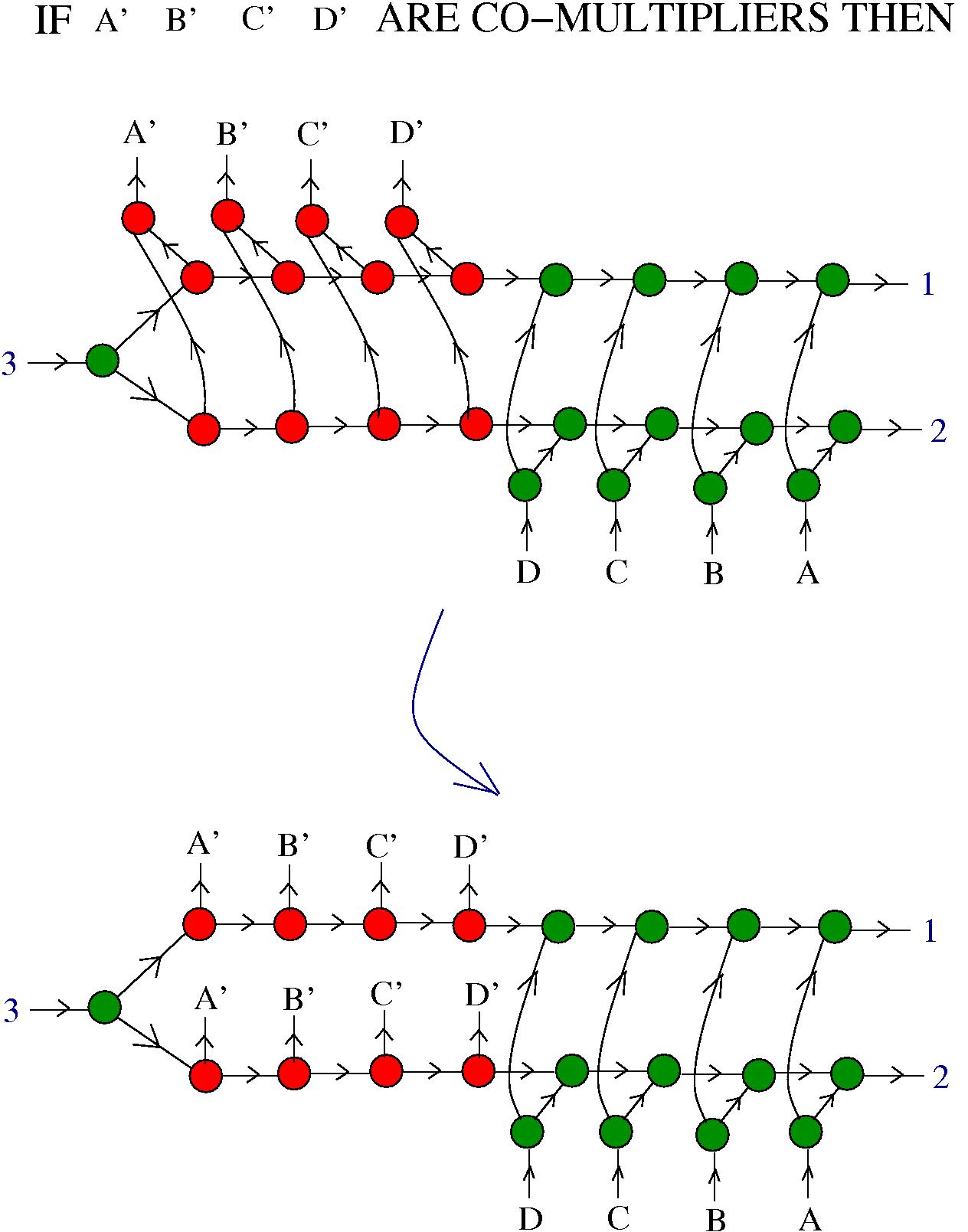}}

In conclusion: if $ A, B, C, D$ are multipliers and $ A', B', C', D'$ are co-multipliers, then the zipper is a multiplier. This will be used in the next section.

\section{Using the chemical concrete machine (II). Turing universality}

The \href{http://en.wikipedia.org/wiki/B,C,K,W_system}{B,C,K,W system } "is a variant of \href{http://en.wikipedia.org/wiki/Combinatory_logic}{combinatory logic} that takes as primitive the combinators $B, C, K$, and $W$. This system was discovered by \href{http://en.wikipedia.org/wiki/Haskell_Curry}{Haskell Curry} in his doctoral thesis". 

I shall explain first which are the correspondents of the B, C, K, W, combinators in the formalism of the chemical concrete machine.  (Via the red-green vs black-white change of notation, they can be deduced from their expressions in graphic lambda calculus, which are obtained by using the algorithm described in section 3 \cite{bgraph}. )

\vspace{.5cm}

\centerline{\includegraphics[width=  80mm]{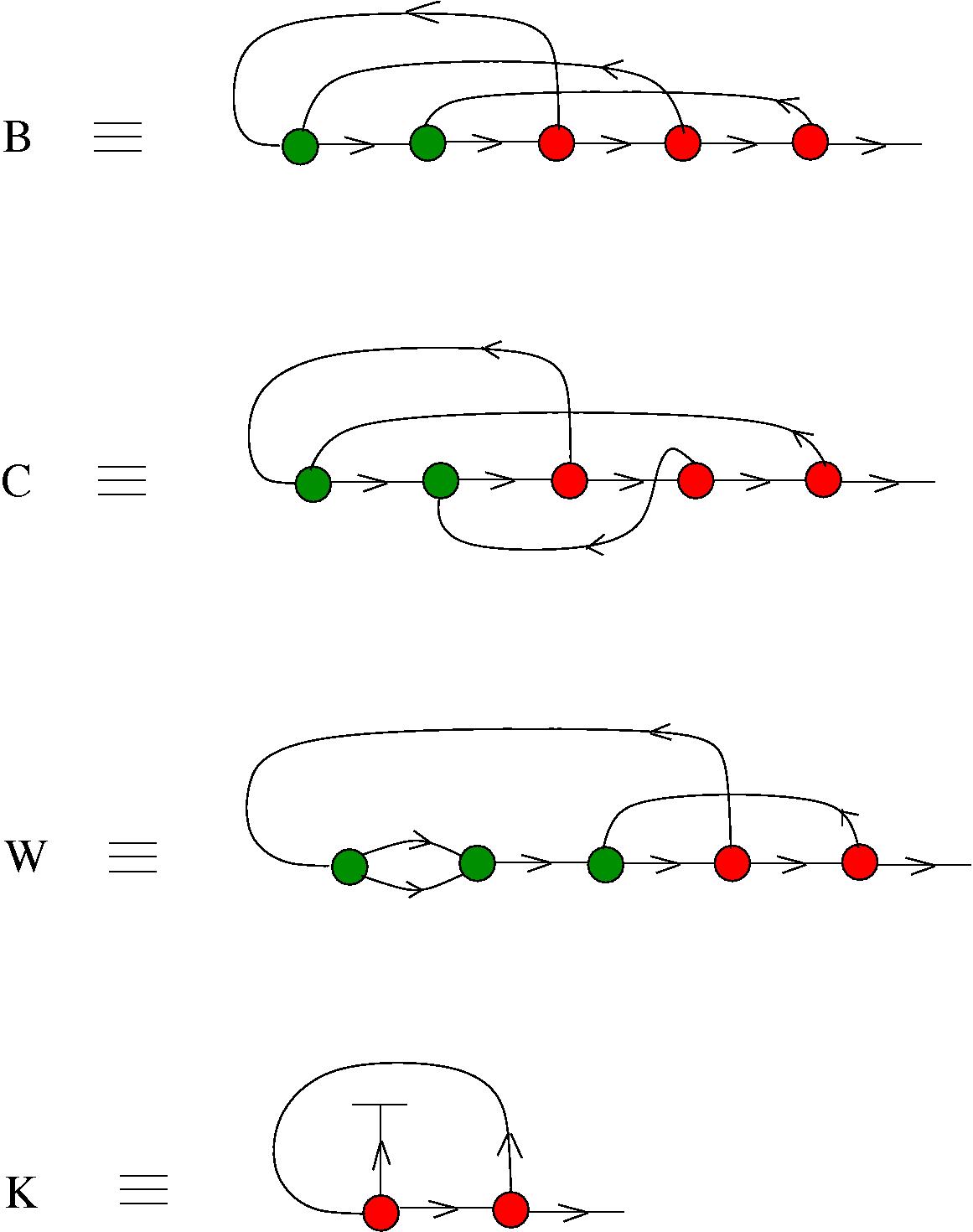}}

\vspace{.5cm}

\begin{definition} The class of combinator molecules (i.e. those which correspond to combinators written in the B,C,K,W system) is defined by the following procedure: 
\begin{enumerate}
\item[-] the B, C, K, W molecules are combinator molecules, 
\item[-] if A, D are combinator molecules then the molecule AD, defined as in the next figure, is a combinator molecule, 
\item[-] if A is a combinator molecule and B is another molecule obtained from A by the application of the moves of the chemical concrete machine, then B is a combinator molecule.  
\end{enumerate}

\vspace{.5cm}

\centerline{\includegraphics[width=  50mm]{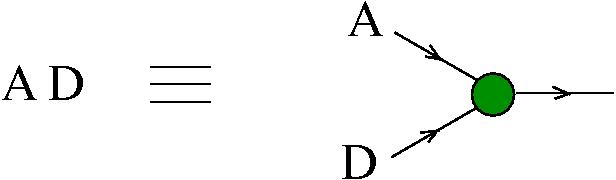}}

\vspace{.5cm}

\label{defcomb}
\end{definition}

The purpose of this section is to prove the following. 

\begin{theorem}
The chemical concrete machine contains  combinatory logic under the form of the B,C,K,W system. 
\label{th1}
\end{theorem}

\paragraph{Proof.} By Theorem 3.1 and Proposition 3.2 \cite{bgraph}, using also the dictionary between the notations in graphic lambda calculus and those of the chemical concrete machine, it follows that the combinatory algebra (in particular in the form of the B,C,K,W system) can be expressed in the chemical concrete machine if: 
\begin{enumerate}
\item[-] we replace the GLOBAL PRUNING move from graphic lambda calculus with the management of the GARB (garbage) class of molecules (alternatively, we might simply neglect GLOBAL PRUNING and we obtain something a bit more expressive than combinatory algebra), 
\item[-] we add to the chemical machine list of moves the following GLOBAL FAN-OUT move: 
\end{enumerate}

\vspace{.5cm}

\centerline{\includegraphics[width=  100mm]{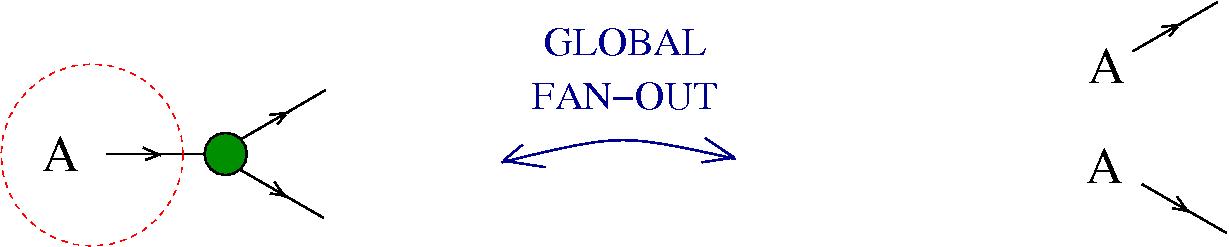}}

\vspace{.5cm}

Here, the meaning of the red dashed circle is that the GLOBAL FAN-OUT move can be applied only if the molecule A has no arrows outside the red dashed circle, with the exception of the one connecting it with the fan-out molecule. 

In Proposition 3.2 \cite{bgraph} was used the S,K,I combinators system, instead of the B, C, K, W, but the Theorem 3.1 \cite{bgraph} shows that already untyped lambda calculus can be expressed with the chemical concrete machine, provided that the GLOBAL FAN-OUT move is added. The choice of the B,C,K, W system instead of the S,K,I system is only for exposition purposes. 

The goal is to show that for any combinator molecule, the GLOBAL FAN-OUT move can be replaces by a finite chain of local moves of the chemical concrete machine. As a consequence, we don't need the GLOBAL FAN-OUT move in order to have combinatory logic in the chemical concrete machine. 

By looking at the last part of the previous section, we want to prove that any combinator molecule is a multiplier. This is done in the following two steps. 

{\bf Step 1.} If A, D are combinator molecules which are multipliers then AD is a multiplier. The proof is given in the next figure, by using DIST moves: 

\vspace{.5cm}

\centerline{\includegraphics[width=  70mm]{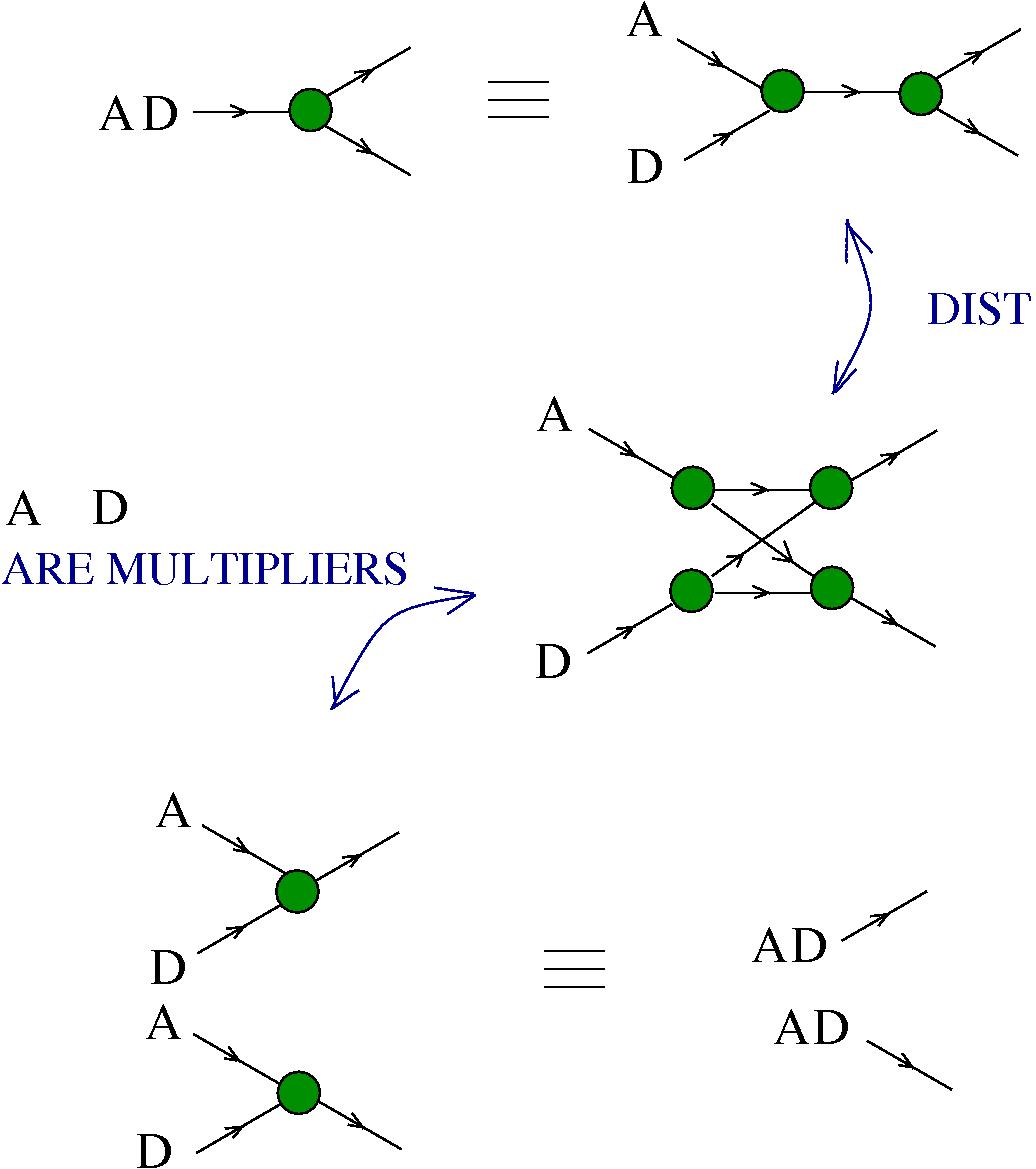}}

\vspace{.5cm}

{\bf Step 2.} Is only left to prove that B,C, K, W are multipliers.  By the conventions of the chemical concrete machine, I mention here the enzymes which are involved in the reactions, instead of writing the moves, like in the graphic lambda calculus.

 The proofs for B and C are very much alike, therefore I put here only the proof that B is a multiplier:

\vspace{.5cm}

\centerline{\includegraphics[width=  100mm]{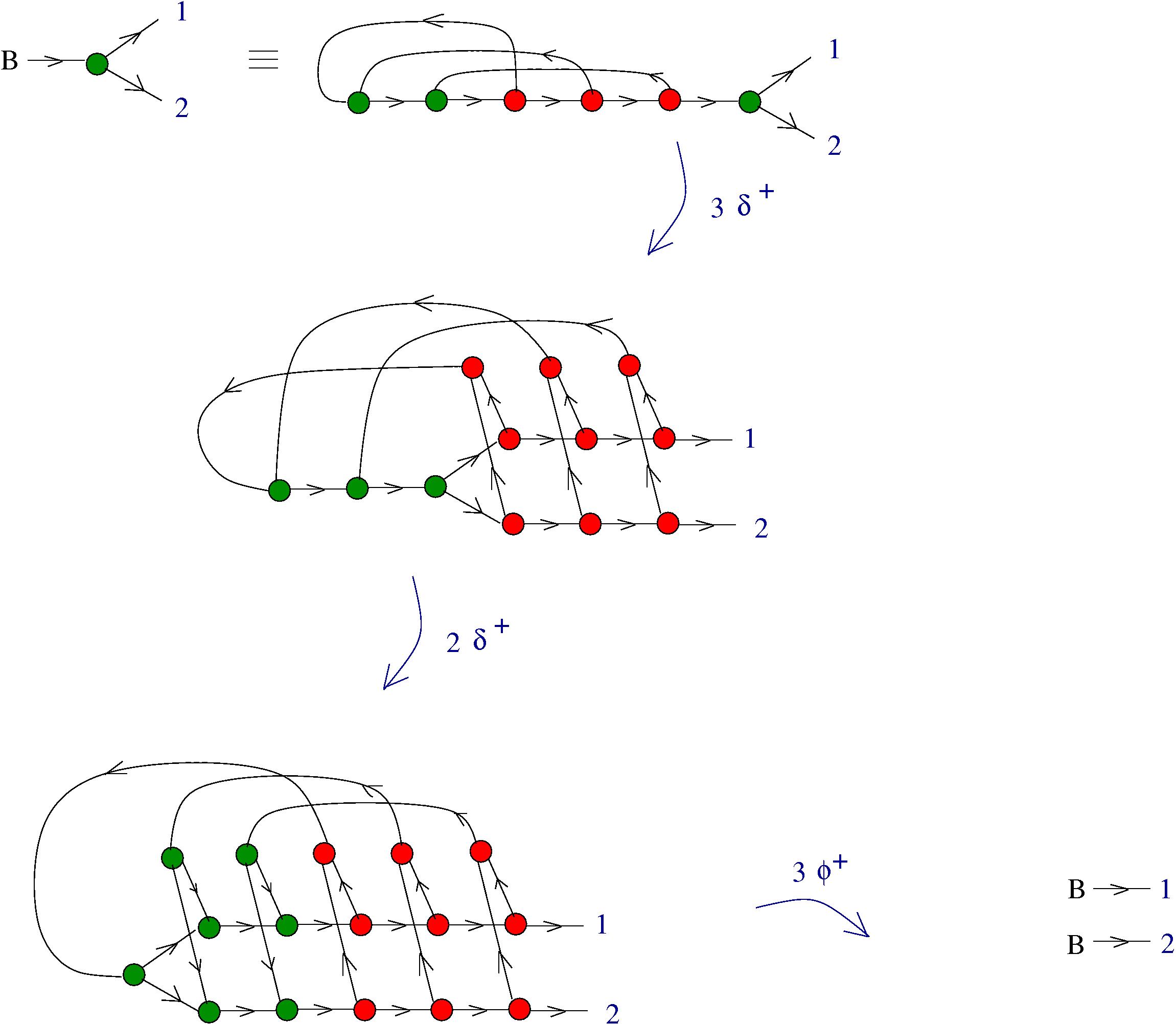}}

\vspace{.5cm}

The proof that K is a multiplier is the following: 

\vspace{.5cm}

\centerline{\includegraphics[width=  90mm]{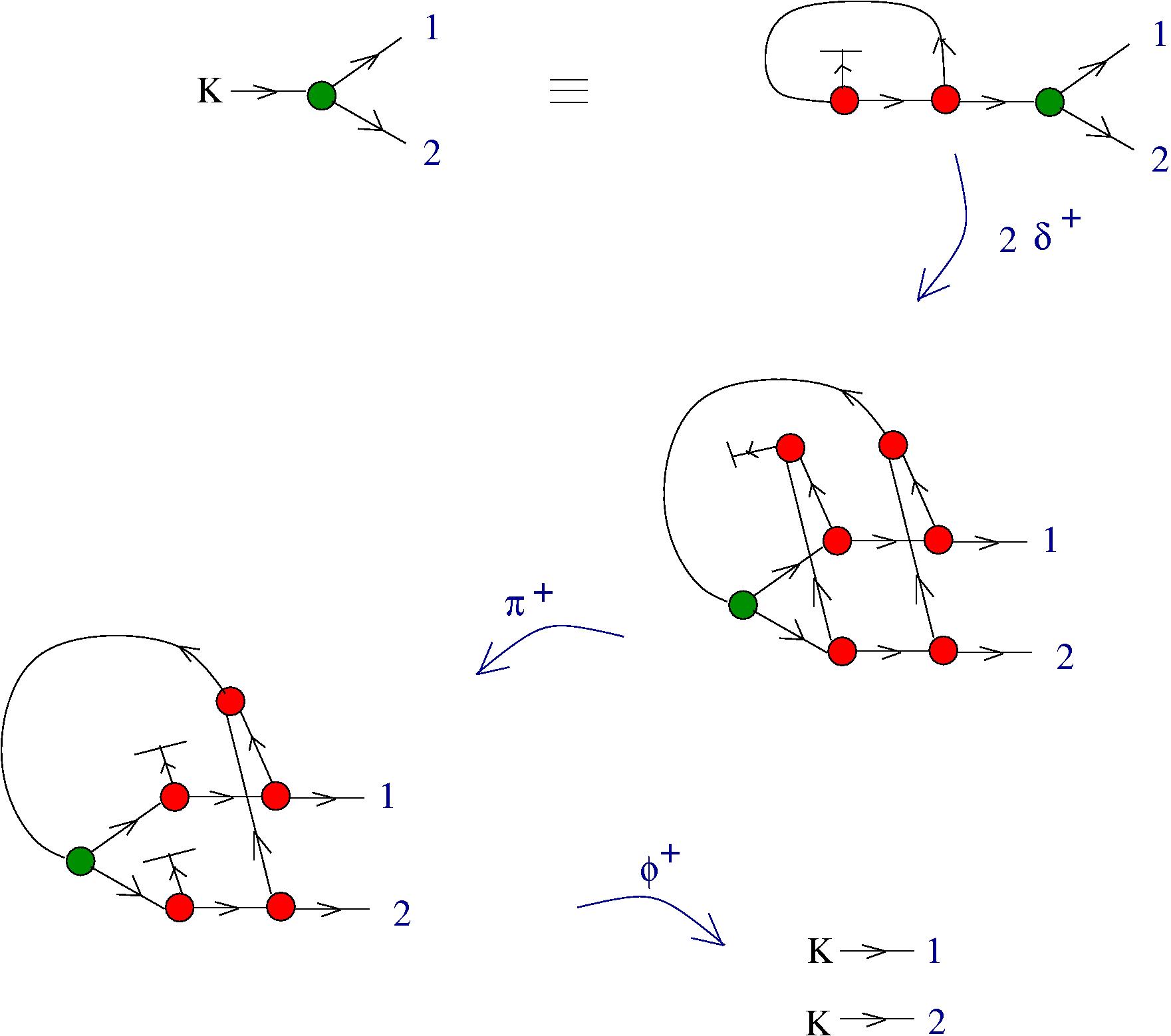}}
 
\vspace{.5cm}

As a side remark, notice how, in both cases, the reactions seem feasible, in the sense that  they can be accomplished by a linear process, because at any step there is only one kind of reaction site available. 

For the W combinator (molecule), things get a bit more complex. 

\vspace{.5cm}

\centerline{\includegraphics[width=  90mm]{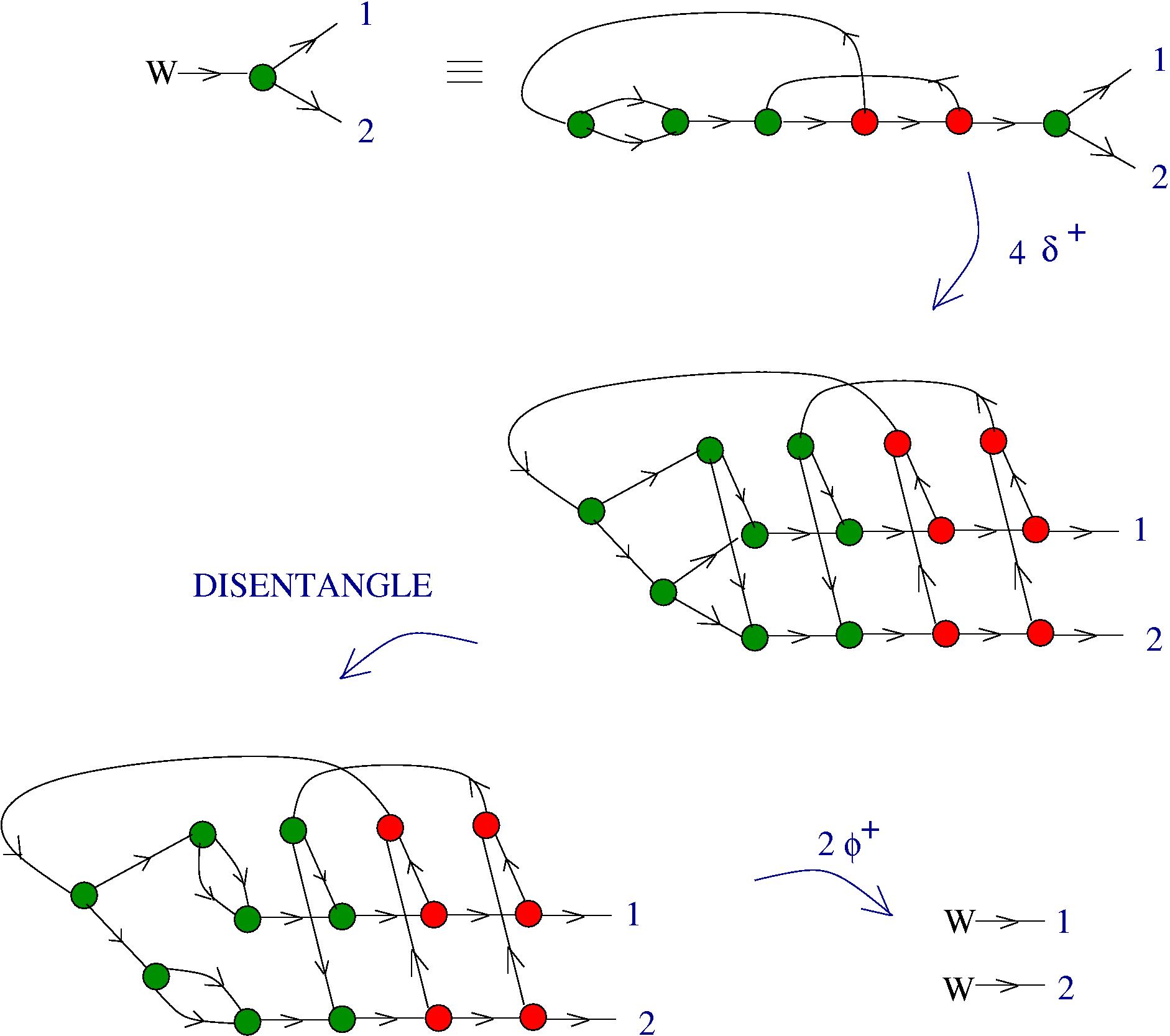}}

\vspace{.5cm}

There is a reaction (or move) which needs explanations. I called it DISENTANGLE (CO-ASSOC) reaction. It is this:

\vspace{.5cm}

\centerline{\includegraphics[width=  80mm]{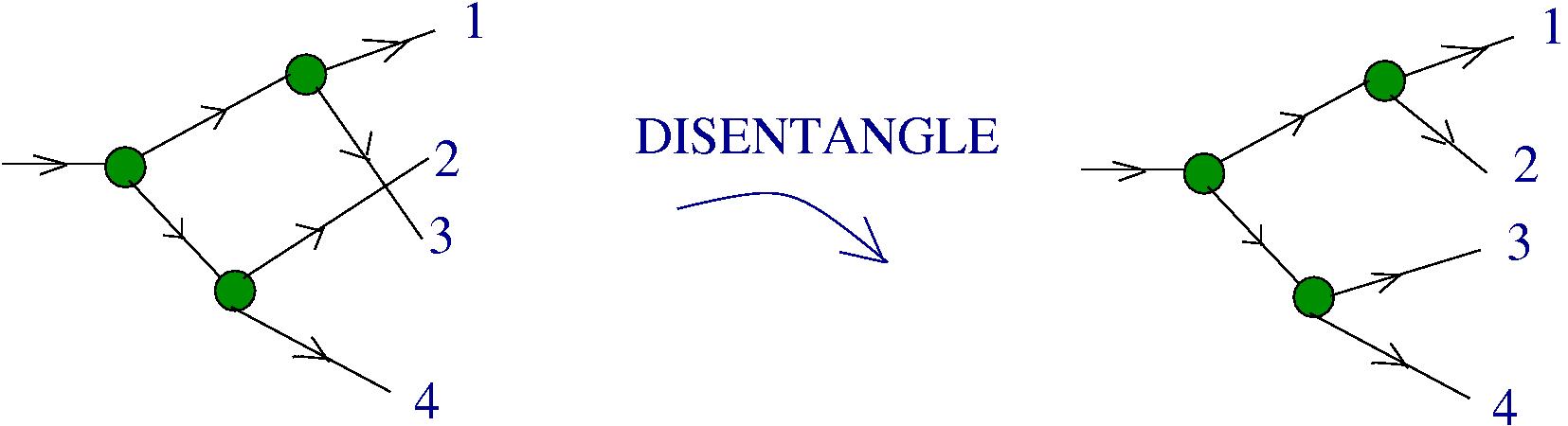}}

\vspace{.5cm}

It can clearly be done by a  succession of CO-ASSOC moves (reactions).  This ends the proof of the theorem. \hfill $\square$

\vspace{.5cm}

From the point of view of the feasibility in the real world (provided a real implementation of the chemical concrete machine will appear), it seems hard to control the exact order of applications of CO-ASSOC moves which gives the DISENTANGLE move as an effect. So, probably,  we shall need a "disentangle enzyme" dedicated to this.

As an alternative, remark that for proving that W is a multiplier we need an application of the DISENTANGLE composite move, described in the next figure:

\vspace{.5cm}

\centerline{\includegraphics[width=  90mm]{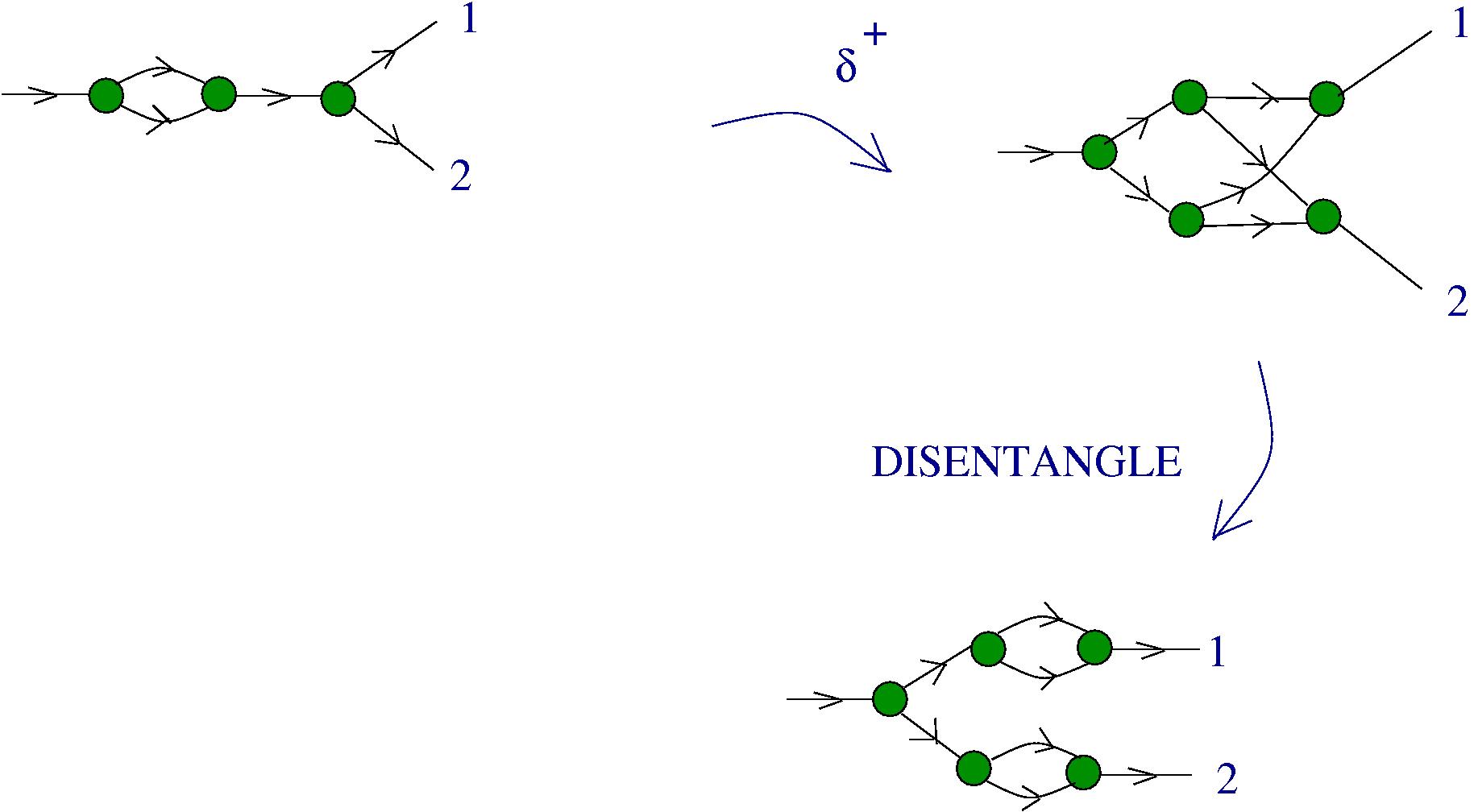}}

\vspace{.5cm}

For practical (or theoretical as well) purposes, it is enough to take this as a move. 
In other words, what would get rid of needing a controlled sequence of CO-ASSOC reactions for multiplying the molecule W is this:  assume that the molecule which is connected to the "y" essential molecule (i.e. to the input of a FAN-OUT gate) is a "propagator". Propagators are defined in the next figure:

\vspace{.5cm}

\centerline{\includegraphics[width=  90mm]{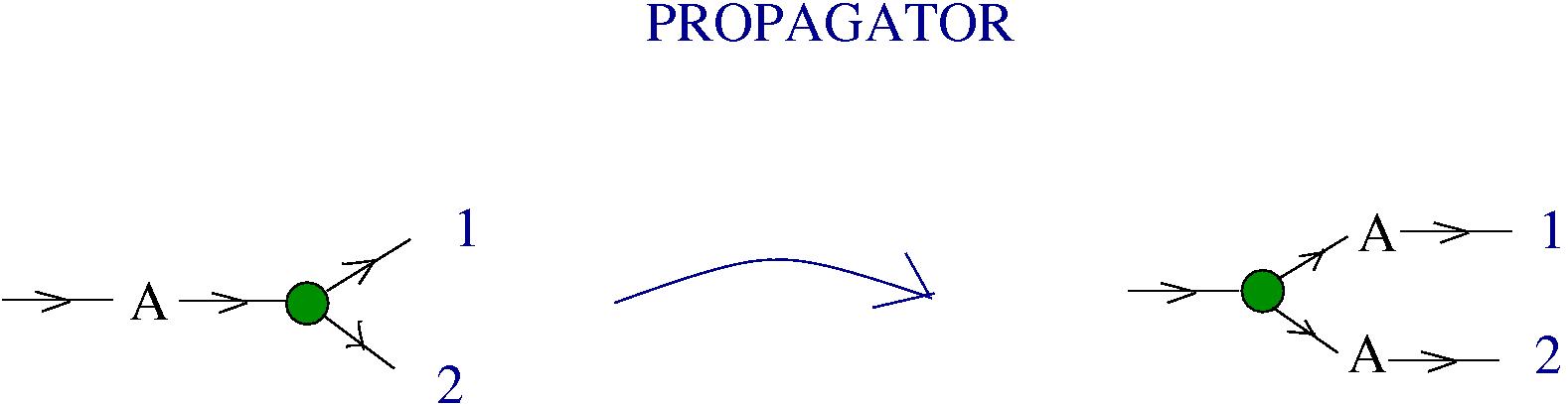}}

\vspace{.5cm}

Propagators are different from multipliers, because they are molecules with one selected input and one selected output which "propagate along a FAN-OUT gate", while multipliers are multiplied by a FAN-OUT gate. Propagators can serve in fact as labels, or names, which propagate along a tree of FAN-OUT gates.

Let's see, as an application, how the IFTHENELSE construct appear as a combinator molecule in the chemical concrete machine.  In lambda calculus, there are terms called TRUE, FALSE and IFTHENELSE, which are the  of the booleans true, false and if-then-else. The associated graphs in the chemical concrete machine are:

\vspace{.5cm}

\centerline{\includegraphics[width=  90mm]{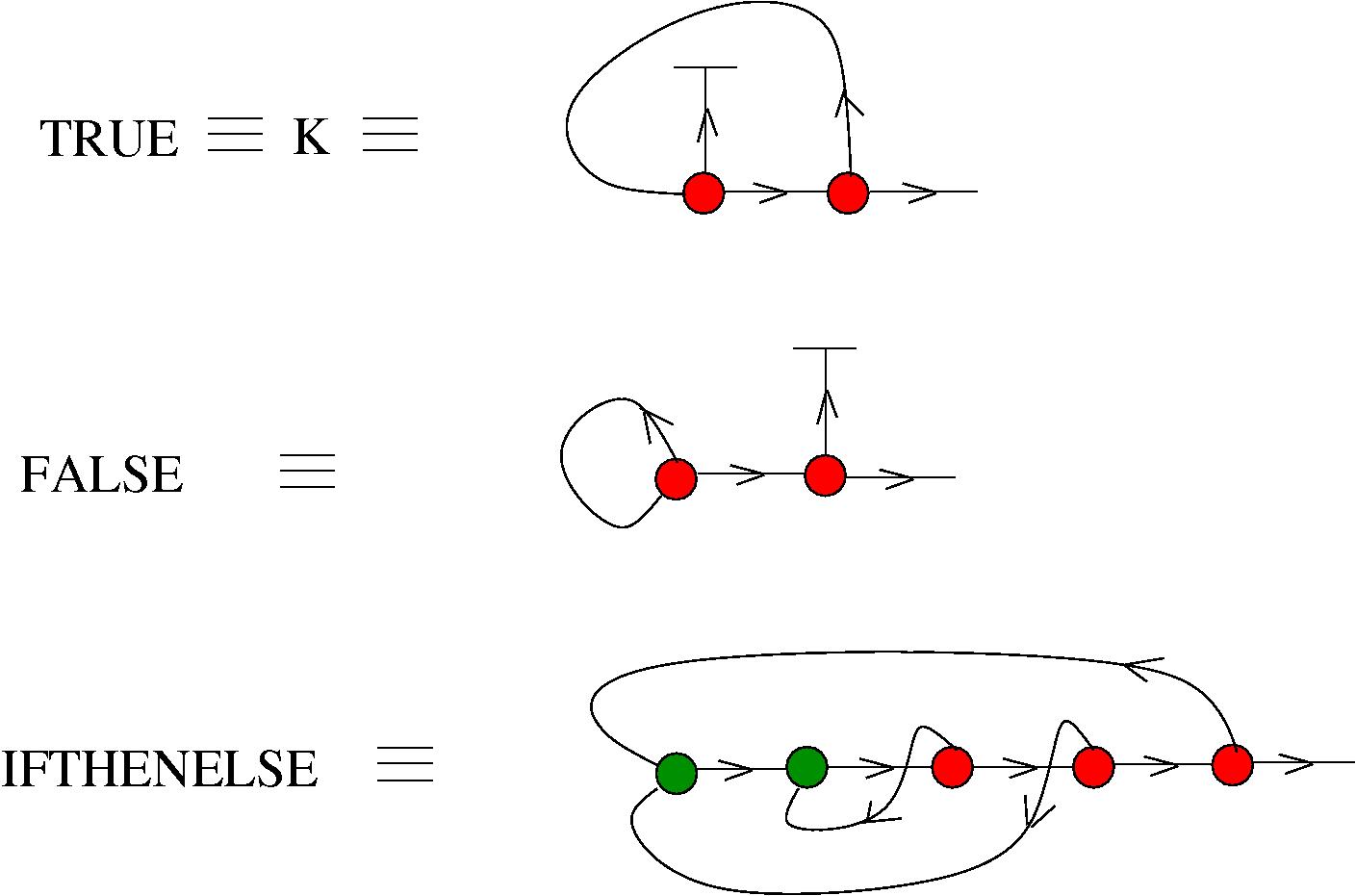}} 

\vspace{.5cm}

Take  two molecules A, B, with one exit each. Then IFTHENELSE TRUE A B should become A. In the chemical concrete machine, with only $\beta^{+}$ enzymes, the following chain of reactions happens: 

\vspace{.5cm}

\centerline{\includegraphics[width=  100mm]{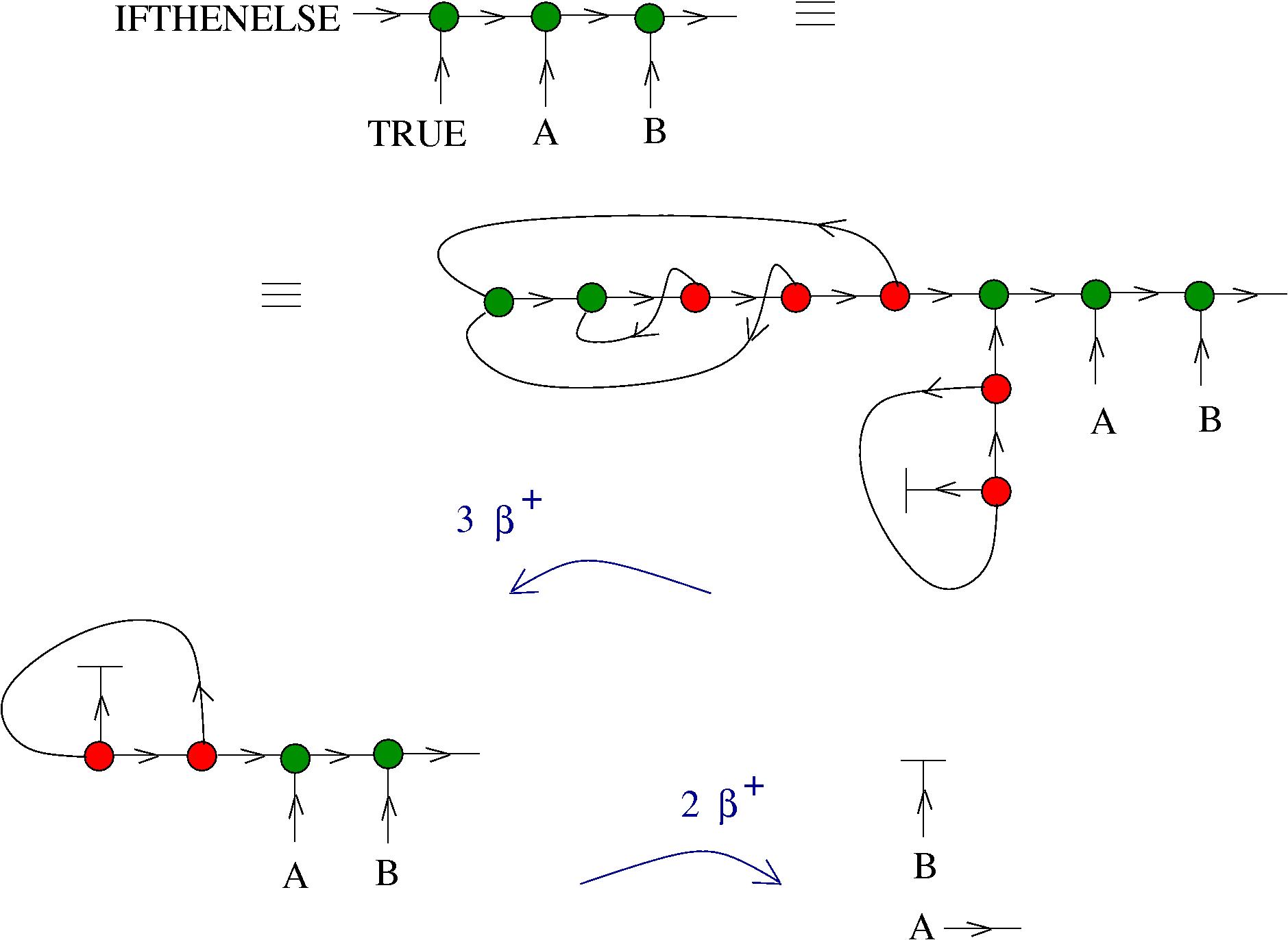}} 

\vspace{.5cm}

Along this chain of reactions, there is no other choice than the ones from the figure, because at every step there is only one reaction site available to the enzyme $\beta^{+}$. The result is, unsurprisingly, compatible with the lambda calculus version, with the exception that A and B are not supposed to be (graphs corresponding to) lambda terms. They can be anything,  for example, from the family of "other molecules".

In lambda calculus IFTHENELSE FALSE A B should become (by reductions) B. In the chemical concrete machine the following chain of reactions happens: 

\vspace{.5cm}

\centerline{\includegraphics[width=  100mm]{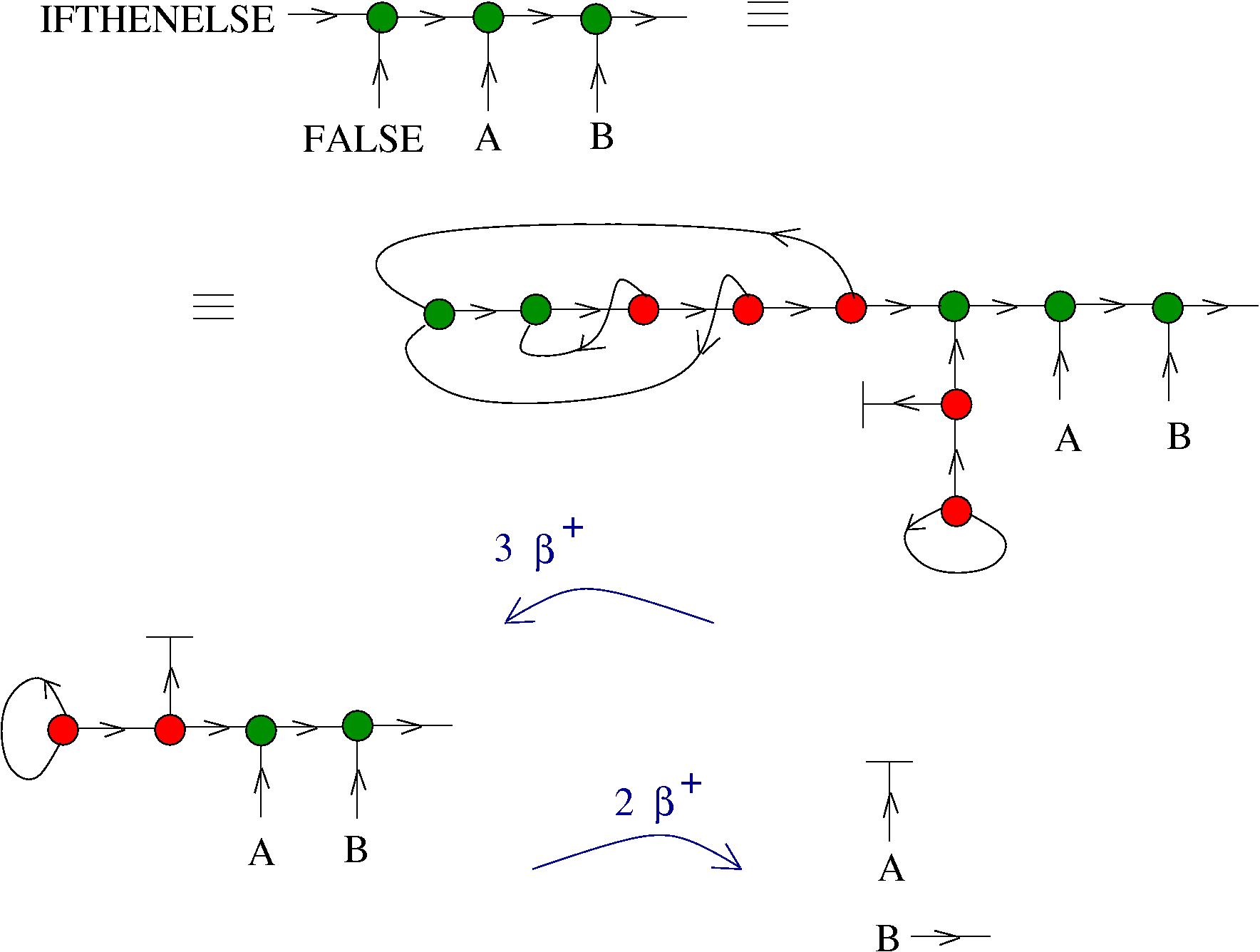}} 

\vspace{.5cm}

With a little bit of imagination, if we look closer to what TRUE, FALSE and IFTHENELSE are doing, we see that it is possible to  adapt the IFTHENELSE to a molecule which releases, under the detection of one  molecule (like TRUE), the "medicine" A, and under the detection of another molecule (like FALSE)  the "medicine" B.

\end{document}